\documentclass[manuscript]{aastex631}

\usepackage{amsmath}
\usepackage{chemformula}
\let\tablenum\relax
\usepackage{siunitx}
\usepackage[verbose]{placeins}
\usepackage{graphicx}
\usepackage{chngpage}
\usepackage{xr}
\usepackage{natbib}
\setcitestyle{numbers}

\begin{document}

\title{Three-stage Formation of Cap Carbonates after Marinoan Snowball Glaciation Consistent with Depositional Timescales and Geochemistry}

\author[0000-0003-2457-2890]{Trent B. Thomas}
\affiliation{Department of Earth and Space Sciences, University of Washington, Seattle, WA, USA}
\affiliation{Astrobiology Program, University of Washington, Seattle, WA, USA}

\author[0000-0001-5646-120X]{David C. Catling}
\affiliation{Department of Earth and Space Sciences, University of Washington, Seattle, WA, USA}
\affiliation{Astrobiology Program, University of Washington, Seattle, WA, USA}

\begin{abstract}

At least two global “Snowball Earth” glaciations occurred during the Neoproterozoic Era (1000-538.8 million years ago). Post-glacial surface environments during this time are recorded in cap carbonates: layers of limestone or dolostone that directly overlie glacial deposits. Postulated environmental conditions that created the cap carbonates lack consensus largely because single hypotheses fail to explain the cap carbonates’ global mass, depositional timescales, and geochemistry of parent waters. Here, we present a global geologic carbon cycle model before, during, and after the second glaciation (i.e. the Marinoan) that explains cap carbonate characteristics. We find a three-stage process for cap carbonate formation: (1) low-temperature seafloor weathering during glaciation generates deep-sea alkalinity;  (2) vigorous post-glacial continental weathering supplies alkalinity to a carbonate-saturated freshwater layer, rapidly precipitating cap carbonates; (3) mixing of post-glacial meltwater with deep-sea alkalinity prolongs cap carbonate deposition. We suggest how future geochemical data and modeling refinements could further assess our hypothesis.

\end{abstract}

\section{Introduction} \label{intro}


Earth's Neoproterozoic Era (1 billion years ago to 538.8 million years ago) is marked by dramatic global climate change. Geologic evidence indicates two major glacial intervals where ice sheets reached low latitudes for millions of years \citep[e.g., reviewed by][]{hoffman_snowball_2017}. These ``snowball Earth'' events \citep{kirschvink_late_1992,hoffman_neoproterozoic_1998} are the Sturtian from 717 to 659 million years ago (Ma) and the Marinoan from ca 645 to 635 Ma. Together, these events bookend the Cryogenian Period (720-635 Ma).

The Cryogenian glacial intervals occurred alongside other global changes, including the appearance of the first large, complex organisms in Earth's history in the subsequent Ediacaran Period \citep{narbonne_ediacara_2005}, an increase of atmospheric \ch{O2} relative to low levels in the mid-Proterozoic \citep{Planavsky.2018}, large excursions in the global carbon isotope record \citep{kaufman_neoproterozoic_1995}, and the break-up of Rodinia and later assembly of Gondwana supercontinents \citep{cordani_rodinia_2003}. All of these transitions occurred during or continued after the Cryogenian Period, which is only $\sim2\%$ of Earth's 4.5-billion year history. Despite this temporal connection, the causes and the relationships between these global changes remain unclear.

Cap carbonates (CCs) probe Earth's surface environment during, and immediately after, the Cryogenian. CCs are layers of limestone or dolostone up to $\sim200$ meters thick that sharply overlie Sturtian and Marinoan glacial deposits in over 50 locations on all major Neoproterozoic continents \citep[reviewed by][]{yu_cryogenian_2020}. Carbonates record Earth's surface conditions because they are sensitive to the chemistry of the atmosphere and ocean from which they precipitate. Thus, the sharp distinction between the glacial deposits and CCs is interpreted as an abrupt shift in Earth's surface environment at the end of each Cryogenian glacial interval from cold, frozen conditions to hot conditions with a high partial pressure of atmospheric carbon dioxide, \ch{pCO2} \citep{hoffman_neoproterozoic_1998}. In this scenario, continental weathering was inhibited during the glacial intervals, allowing volcanic \ch{CO2} to accumulate in the atmosphere and provide enough greenhouse warming to overcome the high albedo of Earth's ice-covered surface, causing deglaciation. The post-glacial Earth then entered a high \ch{pCO2} but low albedo state in which continental weathering produced cations and carbonate ions (i.e. alkalinity) that drove rapid carbonate deposition \citep{kirschvink_late_1992, hoffman_neoproterozoic_1998}. 

The above explanation for CC deposition is broadly consistent with the geologic evidence, but a complete explanation must answer several key questions. First, was the alkalinity source from continental weathering sufficient for the CCs? It is estimated that the global mass of the Marinoan CCs is over $10^{18}$ kg \citep{yu_cryogenian_2020}. Enough alkalinity to generate this much carbonate must be supplied after the glacial interval. Second, what was the timescale of CC deposition? The interpretation of sedimentary structures, the presence of paleomagnetic reversals, and radiometric dating yield conflicting estimates that have yet to be resolved (Table \ref{timescales}). Third, what were the physical and chemical properties of the water body from which the CCs precipitated? After the Marinoan glaciation, the post-glacial ocean was subject to a large influx of glacial meltwater, sea level rise, and transgression onto the land. The deposition of the CCs was likely influenced by these changing ocean conditions.

Many explanations for CC deposition have been proposed, but none are complete. As summarized in \citet{yu_cryogenian_2020}, suggestions include oceanic overturn \citep{grotzinger_anomalous_1995,knoll_comparative_1996}, continental weathering \citep{hoffman_snowball_2002,hoffman_snowball_2017}, gas hydrate destabilization \citep{kennedy_are_2001,kennedy_snowball_2008}, glacial meltwater plumes and subsequent ocean overturn \citep{shields_neoproterozoic_2005}, sediment starvation \citep{Kennedy.2011,spence_sedimentological_2016}, microbial activity \citep{nedelec_sedimentology_2007,font_fast_2010}, and calcareous loess \citep{retallack_neoproterozoic_2011}. Also summarized in \citet{yu_cryogenian_2020}, these explanations all have unresolved deficiencies related to the physical and chemical conditions of the post-glacial ocean, the interpretation of the geologic evidence, and the predicted timescale of deposition.

Advances in our knowledge of the geologic carbon cycle and of Cryogenian conditions allow new tests of hypotheses for CC deposition.

Seafloor weathering has been recognized as a process in the geologic carbon cycle that has been important during some times in Earth's history. Seafloor weathering occurs when seawater circulating through oceanic crust at low-temperatures reacts with constituents of basaltic rock (e.g., volcanic glass, olivine, and plagioclase) to release alkalinity in the form of Ca ions \citep[e.g. reviewed by][]{coogan_low-temperature_2018}. \citet{krissansen-totton_constraining_2017} developed an empirically justified parameterization of seafloor weathering in a geologic carbon cycle model, and \citet{krissansen-totton_constraining_2018} used this model to show that seafloor weathering may have been comparable in strength to continental weathering at some points in Earth's history. The role of seafloor weathering in CC deposition has not been assessed. Previous global geologic carbon cycle models applied to Cryogenian glacial intervals used theoretical rate parameterizations that have proven inconsistent with recent experiments \citep{le_hir_geochemical_2008}, or they omit low-temperature seafloor weathering \citep{penman_coupled_2019,hood_neoproterozoic_2022,fang_coupled_2022}.

\begin{table}[]
    \caption{Estimates for the age and timescale of Marinoan cap carbonate deposition. To calculate depositional timescale from paleomagnetic data, we assume 1 reversal occurred every 250 kyr, consistent with the Miocene, Jurassic, and Cambrian \citep{font_fast_2010}. The sedimentological lines of evidence are broad interpretations of many CC formations across the literature. CC = cap carbonate.}
    \label{timescales}
    \begin{adjustwidth}{-.5in}{-.5in}
    \renewcommand{\arraystretch}{0.8}
    \resizebox{1.05\textwidth}{!}{%
    \begin{tabular}{lllll}
    \hline \hline
    \multicolumn{5}{c}{\textbf{Radiometric Dating}}                                                                                                                                                    \\
    \hline
    \textbf{Age}                    & \textbf{Measurement location}                                 & \textbf{Method} & \textbf{Geologic Formation}                       & \textbf{Reference}         \\
    632.50 $\pm$ 0.48 Ma              & 5 meters above top of CC                                      & U-Pb & Doushantuo Formation, China                       & \citet{condon_u-pb_2005}               \\
    632.3 $\pm$ 5.9 Ma                & 0.9 meters above top of CC                                    & Re-Os           & Sheepbed Formation, Canada                        & \citet{rooney_cryogenian_2015}               \\
    635.23 $\pm$ 0.57 Ma              & Within CC, 2.3 meters above base                                 & U-Pb & Doushantuo Formation, China                       & \citet{condon_u-pb_2005}                \\
    634.57 $\pm$ 0.88 Ma              & Base of CC                                                    & U-Pb & Nantuo Diamictite, China                          & \citet{zhou_calibrating_2019}                 \\
    636.41 $\pm$ 0.45 Ma              & 1 meter below base of CC                                      & U-Pb & Cottons Breccia, Tasmania                         & \citet{calver_globally_2013}               \\
    635.21 $\pm$ 0.59 Ma              & $\sim$30 meters below base of CC                              & U-Pb & Ghuab Formation, Namibia                          & \citet{prave_duration_2016}                \\
    635.5 $\pm$ 1.2 Ma                & $\sim$30 meters below base of CC                              & U-Pb & Ghaub Formation, Namibia                          & \citet{hoffmann_u-pb_2004}              \\
    \hline \hline
    \multicolumn{5}{c}{\textbf{Paleomagnetism}}                                                                                                                                                        \\
    \hline
    \textbf{Depositional Timescale} & \multicolumn{2}{l}{\textbf{Description}}                                        & \textbf{Geologic Formation}                       & \textbf{Reference}         \\
    \textgreater 1.25 Myr           & \multicolumn{2}{l}{5 polarity reversals in first 20 meters of CC}               & Mirassol d'Oeste Section, Brazil                  & \citet{trindade_low-latitude_2003}             \\
    \textgreater 1.25 Myr           & \multicolumn{2}{l}{5 polarity reversals in first 20 meters of CC}               & Terconi Section, Brazil                           & \citet{font_fast_2010}                 \\
    \textgreater 0.5 Myr            & \multicolumn{2}{l}{2 polarity reversals in first 9 meters of CC}                & Jebel Akhdar Section, Oman                        & \citet{kilner_low-latitude_2005}               \\
    \textgreater 0.5 Myr            & \multicolumn{2}{l}{2 polarity reversals in first 12 meters of CC}               & Second Plain Section, Australia                   & \citet{schmidt_palaeomagnetism_2009}              \\
    \hline \hline
    \multicolumn{5}{c}{\textbf{Sedimentology}}                                                                                                                                                         \\
    \hline
    \textbf{Depositional Timescale} & \multicolumn{2}{l}{\textbf{Description}}                       &  \textbf{Geologic Formation}               & \textbf{Review Reference}         \\
    $10^3 - 10^4$ yr                & \multicolumn{2}{l}{Rapid deglaciation and rapid deposition}           &  Many             & \citet{hoffman_snowball_2017}            \\
    \textgreater $10^5$ yr          & \multicolumn{2}{l}{Slow deglaciation and slow deposition}  & Many & \citet{spence_sedimentological_2016}                \\
    \hline \hline
    \end{tabular}
    }
    \end{adjustwidth}
\end{table}

The Marinoan CCs were likely deposited in a stratified, post-glacial ocean. The large volume of glacial meltwater following deglaciation should have created a distinct layer on top of the existing ocean. It has been proposed that the dolostone components of the CCs precipitated out of this layer \citep{shields_neoproterozoic_2005}. This hypothesis is supported by geochemical measurements of the CCs, including $^{87}$Sr/$^{86}$Sr and $\delta^{26}$Mg in multiple formations \citep{liu_geochemical_2013,liu_neoproterozoic_2014,liu_sr_2018} and a global analysis of Ca, Mg, Sr, and C isotopes \citep{ahm_early_2019}. A freshwater layer is consistent with 1D and 3D ocean models, which show that it could last for up to $10^5$ years \citep{yang_persistence_2017,ramme_climate_2022}. Thus, the evidence indicates that the meltwater layer must be considered in CC deposition; however, the previous global geologic carbon cycle models only consider whole-ocean chemistry.

Here, we investigate CC deposition with a geologic carbon cycle model that includes an empirically justified parameterization of seafloor weathering, explicit calculation of chemistry in the post-glacial meltwater layer, and other advances in our knowledge of both the geologic carbon cycle and Cryogenian conditions. Our model is applied to the Marinoan glaciation, but many aspects are likely applicable to the Sturtian and perhaps other glaciations. We identify a mechanism for Marinoan CC deposition that builds on previous explanations to answer the key questions mentioned above, and we find it is consistent with the global collection of CCs (See Supplementary Figure S1).

\section{Results} \label{res}

\subsection{Climate evolution}

\begin{figure}[htbp]
    \centering
    \includegraphics[width=\textwidth]{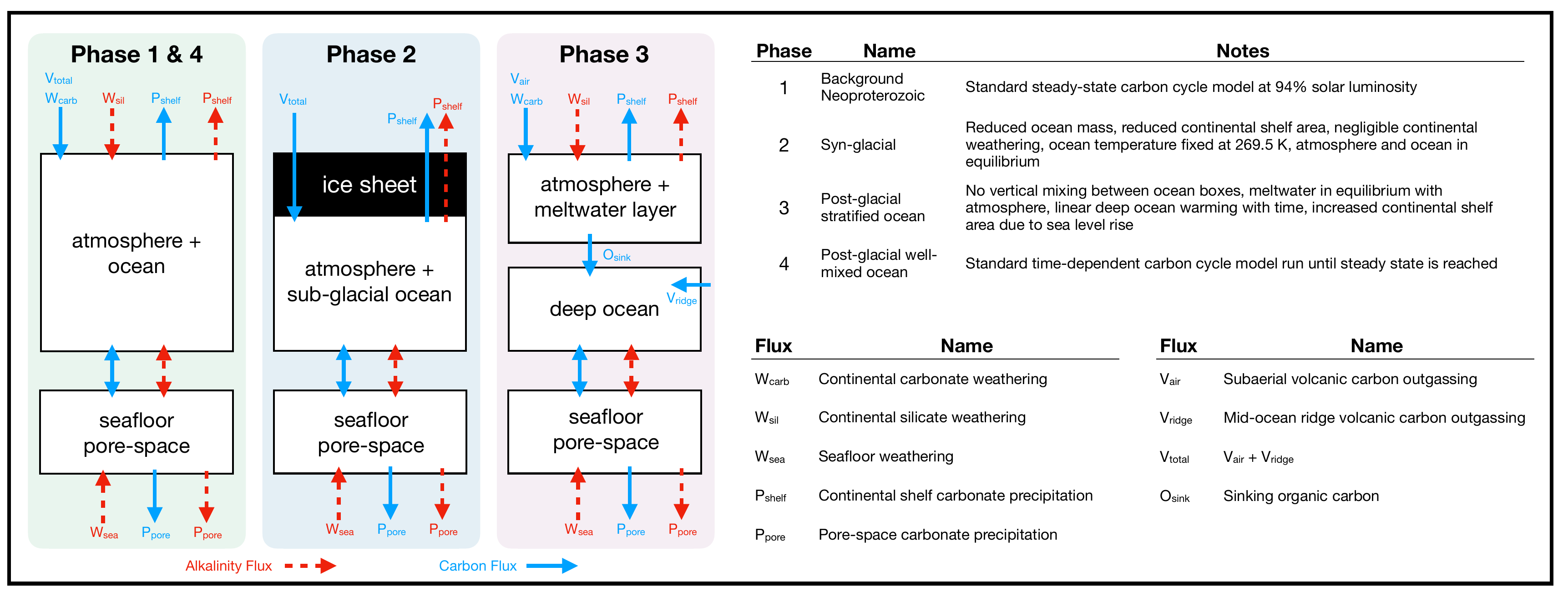}
    \caption{Schematic diagram of the box model used in this work, showing carbon (blue) and alkalinity (red) fluxes in the various model phases.} 
    \label{model_schematic}
\end{figure}

We constructed a model for the evolution of the carbon content and marine alkalinity in Earth's atmosphere and ocean before, during, and after the Marinoan glaciation (Figure \ref{model_schematic}). We use a box model that evolves according to the processes of the geologic carbon cycle, which include continental silicate and carbonate weathering, volcanism, carbonate deposition, and seafloor weathering. In order to capture the unique climates of a glaciation event, we configure the model to address 4 distinct phases: the background Neoproterozoic phase, the syn-glacial phase, the post-glacial stratified ocean phase, and the post-glacial well-mixed ocean phase. 

The basic model builds upon a previously data-validated model \citep{krissansen-totton_constraining_2017, krissansen-totton_constraining_2018} and consists of two boxes: one box for the combined atmosphere-ocean and one box for water in the seafloor rock pore-space, where seafloor weathering takes place. In the post-glacial stratified ocean phase, another box is included to explicitly account for a glacial meltwater layer. After calculating the aqueous chemistry in each box, the model calculates climate variables (e.g., ocean pH, surface temperature, and \ch{pCO2}) and geologic processes (e.g., weathering rates, carbonate deposition rates). A complete description of these calculations is found in Methods and Supplementary Material.

Transitions between model phases are imposed. A rigorous treatment of the transitions would require calculations beyond the scope of this work such as the complex movement of ice sheets subject to the ice-albedo instability or changes in the 3D circulation of the ocean. Thus, model transitions are not explicitly calculated, but they are informed by the literature. We instead focus on the major environmental and chemical conditions within each model phase. 

We sample uncertain model parameters within their plausible value ranges. We follow \citet{krissansen-totton_constraining_2017} and \citet{krissansen-totton_constraining_2018} to sample the key uncertain parameters in the geologic carbon cycle model. We also introduce extra parameters related to the glaciation, such as the size of the ice sheets, the continental shelf area during glaciation, and the composition of glacial meltwater. 

\begin{figure}[htbp]
    \centering
    \includegraphics[width=0.7\textwidth]{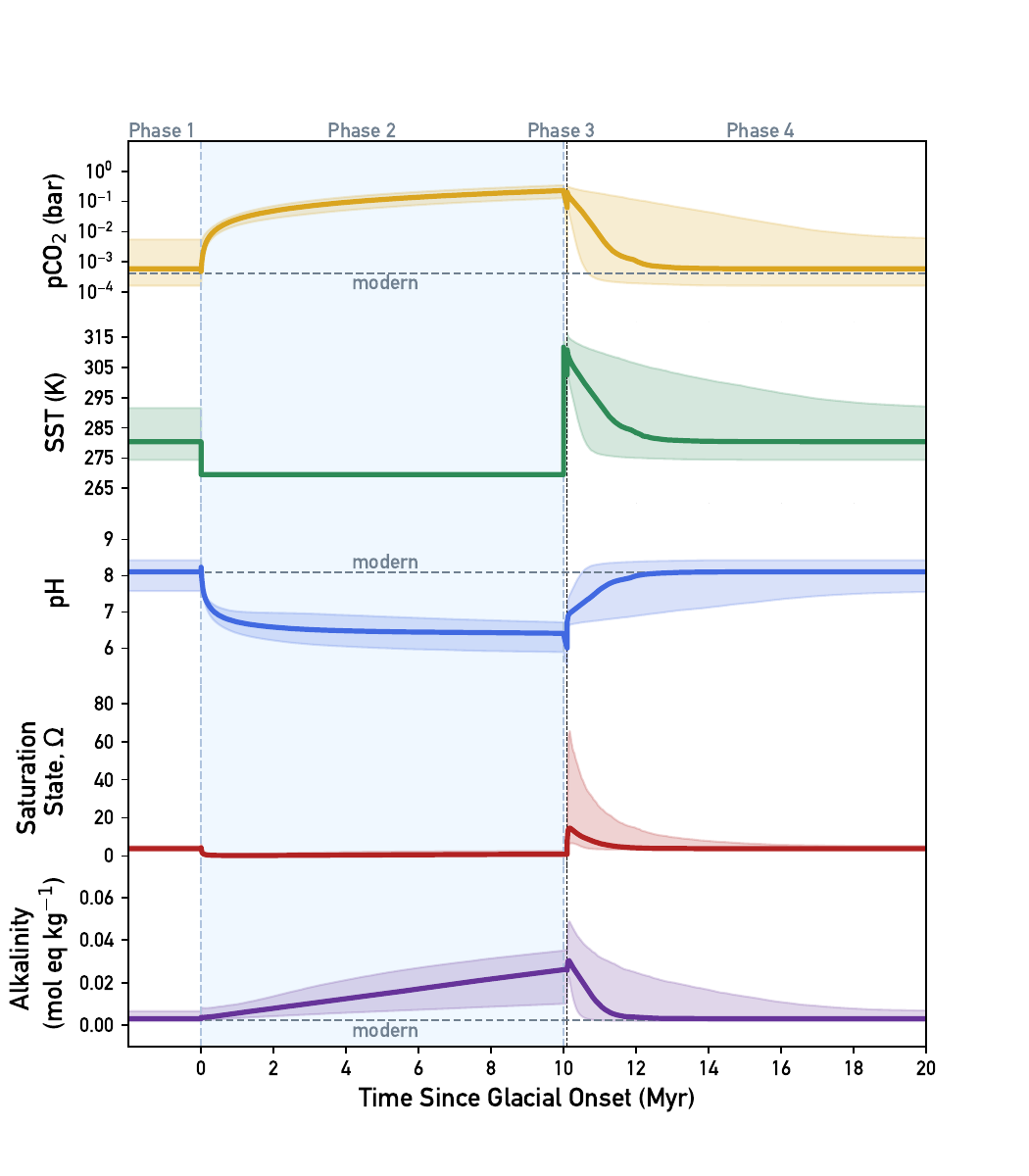}
    \caption{Climate evolution through all modeled phases noting that post-glacial Phase 3 is very compressed on a Myr timescale, so is expanded in Figure \ref{clim_evo_inset}. The shaded regions are the 95\% confidence intervals and the solid lines are the median model runs with randomly sampled parameters. Glaciation was terminated after 10 Myr and the stratified ocean was terminated after 100 kyr ($t_{strat} = 10^5$ years). Vertical dashed lines indicate boundaries between model phases, noting the two close vertical lines at 10 Myr and 10.1 Myr bounding post-glacial stratified ocean Phase 3. Horizontal dashed lines show modern values noted in the main text.}
    \label{clim_evo}
\end{figure}

\begin{figure}[htbp]
    \centering
    \includegraphics[width=\textwidth]{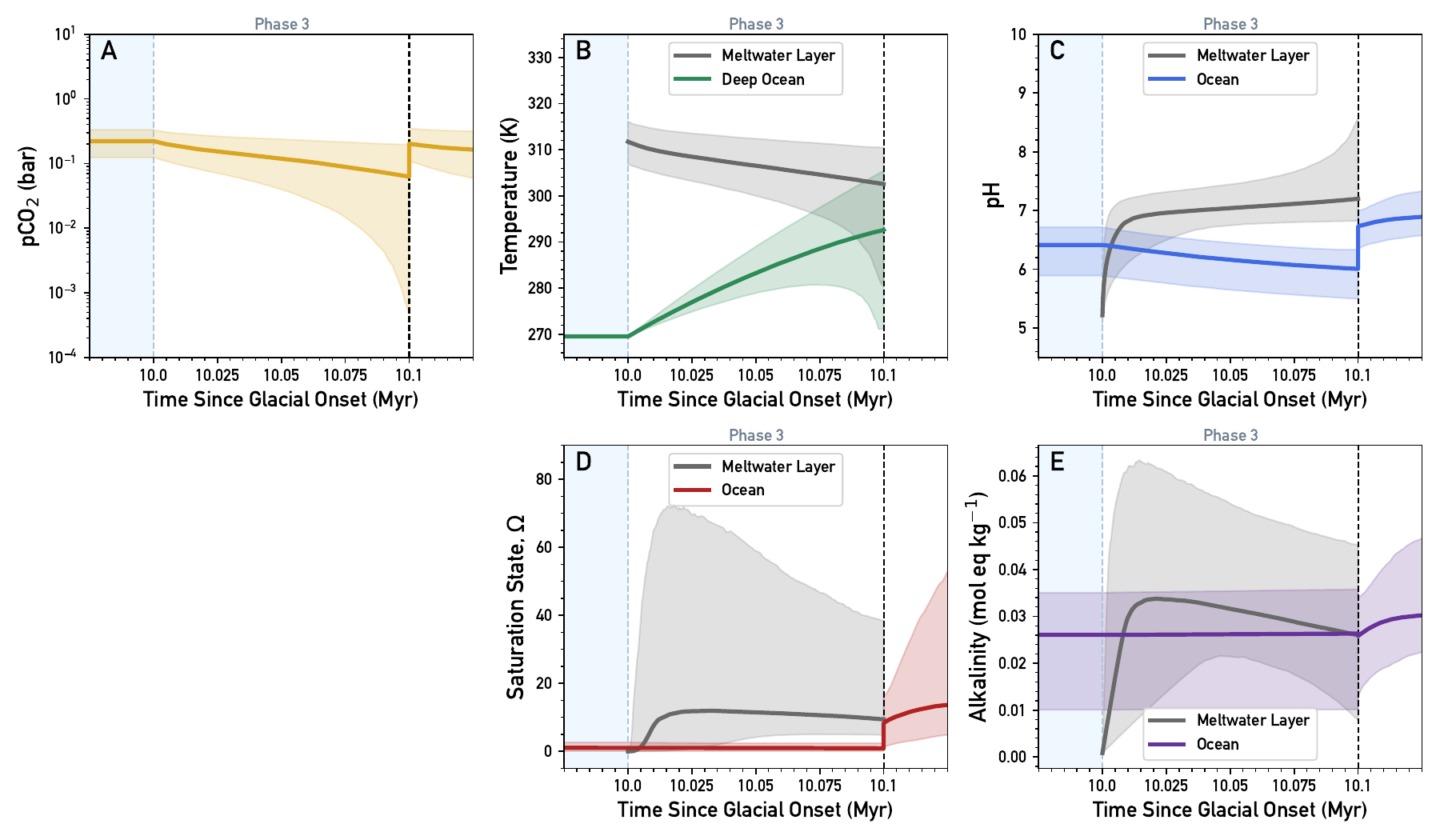}
    \caption{Climate evolution in the stratified ocean Phase 3 from Figure \ref{clim_evo}. The shaded regions are the 95\% confidence intervals and the solid lines are the median model runs with randomly sampled parameters. Post-glacial ocean stratification is assumed to last 100 kyr ($t_{strat} = 10^5$ years), indicated by the vertical dashed black line. Following stratification, the two reservoirs are combined and averaged.}
    \label{clim_evo_inset}
\end{figure}

By modeling the glaciation event in 4 distinct phases, we generate results that are self-consistent across the full range of potential climate states. The nominal model evolution through all phases is shown in Figures \ref{clim_evo} and \ref{clim_evo_inset}. The model is run 3000 times to derive median values and confidence intervals while sampling uncertain parameters.

The pre-glacial phase (Phase 1), given typical Neoproterozoic parameters, is characterized by a temperate or cool background climate. The 95\% confidence interval for equilibrium atmospheric \ch{pCO2} is $\SI{1.6e-4}{}$ to $\SI{5.5e-3}{}$ bar, yielding surface temperatures of $274.4 - 291.6$ K, ocean pH of $7.6 - 8.4$, and ocean calcite saturation state of $3.3 - 5.2$, which is reasonable given uncertainty in marine \ch{Ca^2+} concentration in the Neoproterozoic \citep[e.g.][]{Hill.2000,Arp.2001,Spear.2014}. In the median model evolution, \ch{pCO2} is $\SI{5.9e-4}{}$ bar and the ocean pH is 8.11, which is similar to the modern values of $\SI{4e-4}{}$ bar and 8.1, respectively; however, the reduced solar luminosity allows for lower surface temperatures. The low surface temperature would leave Earth in a state that is potentially vulnerable to a snowball-causing climate perturbation.

During glaciation (Phase 2), \ch{pCO2} steadily rises while seafloor weathering supplies alkalinity to the subglacial ocean. With ice sheets covering the land and a slow hydrologic cycle, we assume that continental weathering either stops completely or is fixed at a rate several orders of magnitude below modern, following \citet{lan_massive_2022}. Seafloor weathering continues though, and causes alkalinity in the sub-glacial ocean to reach up to 0.035 mol eq kg$^{-1}$, which is $\sim14$ times higher than modern seawater \citep[2320 $\mu$mol eq kg$^{-1}$;][p. 131]{pilson_introduction_2012}. The built up alkalinity does not result in widespread carbonate precipitation because high \ch{CO2} levels cause the ocean to be acidic, the ocean is cold, and the thick ice sheets eliminate most of the available shelf area for precipitation (though there is still some carbonate precipitation in the ocean crust). Thus, atmospheric \ch{pCO2} steadily rises as \ch{CO2} is supplied by subaerial and submarine outgassing. Note that we assume the atmosphere and ocean are in equilibrium despite widespread ice sheets, which is further justified in Methods.

We force the glaciation to end after 10 Myr, which is consistent with constraints on the duration of the Marinoan glaciation \citep[see][and references within]{hoffman_snowball_2017}. At the end of glaciation, \ch{pCO2} is 0.13-0.35 bar, consistent with estimates of the required \ch{pCO2} to cause global melting of 0.1-0.3 bar \citep{higgins_aftermath_2003,pierrehumbert_high_2004,hoffman_snowball_2017}. We do not include the complicated process of ice sheet melting in our model, because we focus on the evolution of geochemical parameters.

In the post-glacial stratified ocean (Phase 3), continental weathering rapidly supplies alkalinity to the meltwater layer and CC deposition begins. In the nominal model, we assume that the lifetime of the meltwater layer, $t_{strat}$, is $10^5$ years, but we also consider a scenario with $t_{strat} = 10^4$ years (Supplementary Figures S2 and S3). The equilibration of the meltwater layer with the high \ch{CO2} atmosphere causes immediate and intense acidification. The acidification is counteracted by alkalinity delivered from continental weathering in the hot, high \ch{CO2} conditions. Alkalinity supplied to the meltwater layer can cause the calcium-carbonate saturation state to reach over 70 before carbonate deposition can remove alkalinity at the same rate and balance the system. In the median model runs, carbonate deposition in the meltwater layer removes $0.04$ and $0.16$ bar \ch{CO2} in the first $10^4$ and $10^5$ years of stratification, respectively.

In the post-glacial well-mixed ocean phase (Phase 4), alkalinity is supplied to the surface from the deep ocean, CC deposition continues, and a steady state climate is eventually recovered. As mentioned above, seafloor weathering causes alkalinity to build up in the sub-glacial ocean during Phase 2. During Phase 3, this alkalinity remains trapped in the deep ocean under the meltwater layer. It is not until Phase 4, when the meltwater layer mixes with the underlying ocean, that this alkalinity is delivered to the surface ocean and contributes to carbonate precipitation on continental shelves. Note that in our model construction, the meltwater layer and the underlying ocean are abruptly mixed in a single timestep, which causes the discontinuity seen in some climate variables. This transition was probably more gradual and localized in reality, but we expect the ultimate evolution of the climate variables to be the same. In Phase 4, \ch{CO2} continues to be removed from the atmosphere on the timescale of the geologic carbon cycle, and the background Neoproterozoic steady state climate is generally recovered within 5 Myr of deglaciation.

\subsection{Post-glacial carbonate deposition}

\begin{figure}[htbp]
    \centering
    \includegraphics[width=\textwidth]{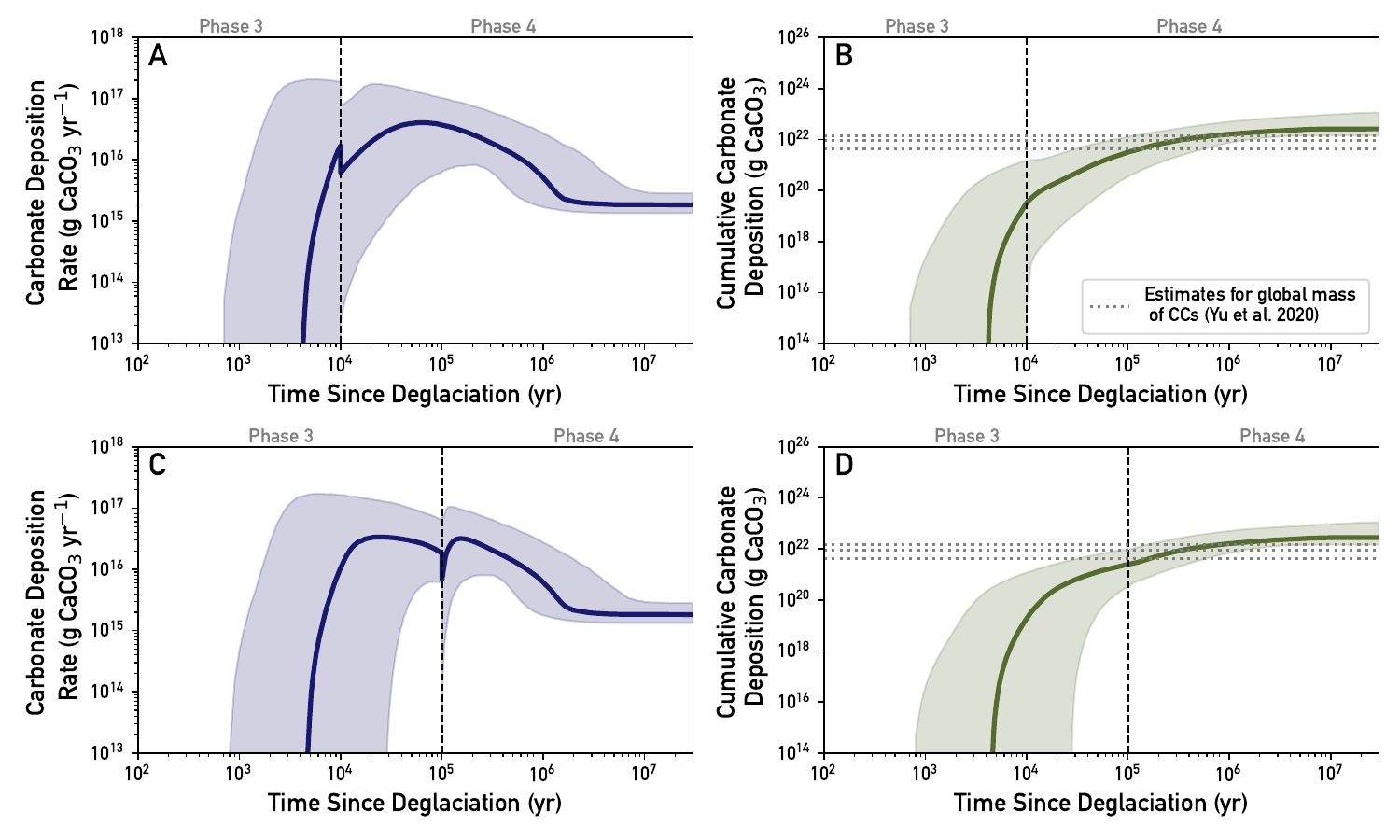}
    \caption{Carbonate deposition rate and cumulative carbonate deposition after deglaciation in terms of mass for different lifetimes of the stratified ocean. The end of the stratified ocean (Phase 3), $t_{strat}$, is indicated by the vertical dashed black lines. In panels \textbf{A} and \textbf{B}, $t_{strat} = 10^4$ years. In panels \textbf{C} and \textbf{D}, $t_{strat} = 10^5$ years. Horizontal dotted lines in panels \textbf{B} and \textbf{D} are minimum, median, and maximum estimates of the global mass of Marinoan cap carbonates from \citet{yu_cryogenian_2020}: $\SI{4.2e21}{}$, $\SI{9.3e21}{}$, and $\SI{14.4e21}{g}$, respectively. The shaded regions are the 95\% confidence intervals and the solid lines are the median model runs with randomly sampled parameters.}
    \label{carbdep_mass}
\end{figure}

\begin{figure}[htbp]
    \centering
    \includegraphics[width=\textwidth]{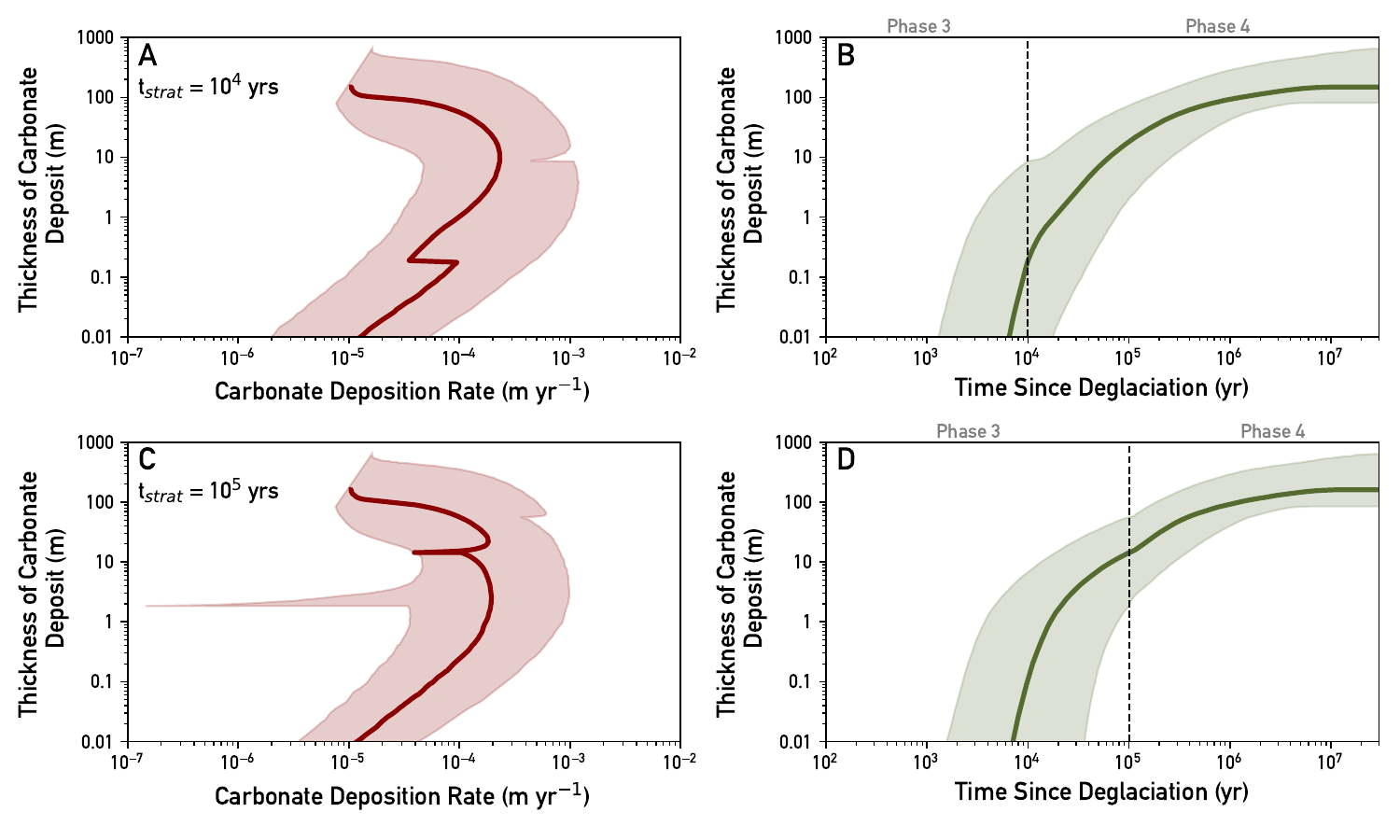}
    \caption{Carbonate deposition rate and cumulative carbonate deposition after deglaciation in terms of deposit thickness for different lifetimes of the stratified ocean. The end of the stratified ocean (Phase 3), $t_{strat}$, is indicated by the vertical dashed black line in panels \textbf{B} and \textbf{D}. In panels \textbf{A} and \textbf{B}, $t_{strat} = 10^4$ years. In panels \textbf{C} and \textbf{D}, $t_{strat} = 10^5$ years. The shaded regions are the 95\% confidence intervals and the solid lines are the median model runs with randomly sampled parameters. Confidence intervals do not align in panels \textbf{A} and \textbf{C} because they are originally time-based but are converted into height-based here.}
    \label{carbdep_height}
\end{figure}

Figures \ref{carbdep_mass} and \ref{carbdep_height} show how carbonates are deposited after the glaciation event in the nominal model runs with $t_{strat} = 10^4$ and $10^5$ years. Confidence intervals are derived from the model runs in Figures \ref{clim_evo} and \ref{clim_evo_inset} and in Supplementary Figures S2 and S3.

The carbonate deposition rate quickly rises in the first 10 kyr after deglaciation and reaches a maximum value 10 to 100 kyr after deglaciation (Figure \ref{carbdep_mass}). For $t_{strat} = 10^4$ years, the median deposition rate peaks $\sim 67$ kyr after deglaciation, when the ocean is well-mixed (Phase 4). For $t_{strat} = 10^5$ years, the median deposition rate peaks $\sim 25$ kyr after deglaciation, when the ocean is still stratified (Phase 3). These results are due to the fact that the meltwater layer can reach a higher saturation state and faster deposition rate when it has a longer lifetime because more alkalinity can be delivered. When $t_{strat} = 10^5$ years, the second peak in deposition rate after the ocean mixes is probably unrealistic because it is due to the abrupt water mixing we impose in the model. In reality, the carbonate deposition rate would likely have been uniformly higher in the stratified ocean phase as the alkalinity from the deep ocean slowly mixed into the meltwater layer.

For context, post-glacial carbonate deposition peaks above $10^{17}$ g \ch{CaCO3} yr$^{-1}$, 2 orders of magnitude higher than modern carbonate deposition on continental shelves \citep[$\sim\SI{1.4e15}{}$ g \ch{CaCO3} yr$^{-1}$;][]{iglesias-rodriguez_progress_2002}. The post-glacial deposition rate is so high primarily because the saturation state in the meltwater layer reaches up to $\Omega = 70$, compared to $\Omega = 3-5$ at the surface of the modern ocean.

From the 95\% confidence intervals for both values of $t_{strat}$, we predict that it took between 32 kyr and 591 kyr to produce the estimated minimum global mass of CCs (Figure \ref{carbdep_mass}). In the median model runs, the time to deposit the CCs is 127 kyr and 162 kyr for $t_{strat} = 10^4$ and $10^5$ yr, respectively. The timescales are similar despite differences in $t_{strat}$ because the sources of alkalinity are the same: seafloor weathering during glaciation and continental weathering after glaciation. The alkalinity sources are the ultimate limiting factor for carbonate precipitation, and the timescale of deposition is set by the time it takes for them to saturate the whole ocean.

The proportion of the CCs deposited from the meltwater layer versus the well-mixed ocean depends on $t_{strat}$ (Figure \ref{carbdep_mass}). For $t_{strat} = 10^4$ years, the median total carbonate mass deposited in the meltwater layer is $\SI{3.3e19}{}$ g, with 95\% confidence interval of $\SI{3.3e15}{}$ g to $\SI{1.4e21}{}$ g. For $t_{strat} = 10^5$ years, the median is $\SI{2.5e21}{}$ g and the 95\% confidence interval is $\SI{3.2e20}{}$ g to $\SI{9.4e21}{}$ g. Most of the CCs are deposited after ocean mixing. Relative to the total carbonate mass deposited in the first 5 Myr after deglaciation, 0.13\% and 9.83\% of the CCs were deposited in the meltwater layer in the median model runs with $t_{strat} = 10^4$ and $10^5$ yr, respectively.

For more direct relevance to the geologic record, we convert the mass of carbonates into a deposit thickness (Figure \ref{carbdep_height}). For a rough comparison, we assume constant sedimentation rates and that the total carbonate mass is evenly spread over a surface area equal to 2 times the modern continental shelf area to account for post-glacial sea level rise (See Methods). This calculation is somewhat qualitative because the dynamics of the depositional environments on post-glacial Earth were complex and subject to regional variability \citep[e.g.,][]{creveling_sea-level_2014,irie_nonmonotonic_2019}. In the median model runs, the peak deposition rate in a globally averaged CC deposit occurs 10.8 m and 2.6 m upsection for $t_{strat} = 10^4$ and $10^5$ yr, respectively. In the 95\% confidence interval for $t_{strat} = 10^4$ years, $<0.001 - 8.12$ m were deposited out of the meltwater layer. For $t_{strat} = 10^5$ years, the range is $1.84 - 53.68$ m.

\subsection{High-alkalinity meltwater scenario}

In the nominal model results, carbonate deposition does not begin until at least $\sim1$ kyr after deglaciation. In this depositional hiatus, continental weathering has not yet supplied enough alkalinity to raise the saturation state of the meltwater high enough for rapid carbonate precipitation. This occurs for two reasons: (1) we assume that the initial alkalinity of the glacial meltwater is low, estimated from modern glacial waters, and (2) we do not explicitly model the global deglaciation, where meltwater from continental ice sheets would have traversed rocky terrain before entering the ocean. In reality, as the Earth was deglaciating, the water from the continental ice sheets would have weathered some of the rock and probably caused the initial alkalinity of the meltwater to be higher than we nominally assume. 

\begin{figure}[htbp]
    \centering
    \includegraphics[width=\textwidth]{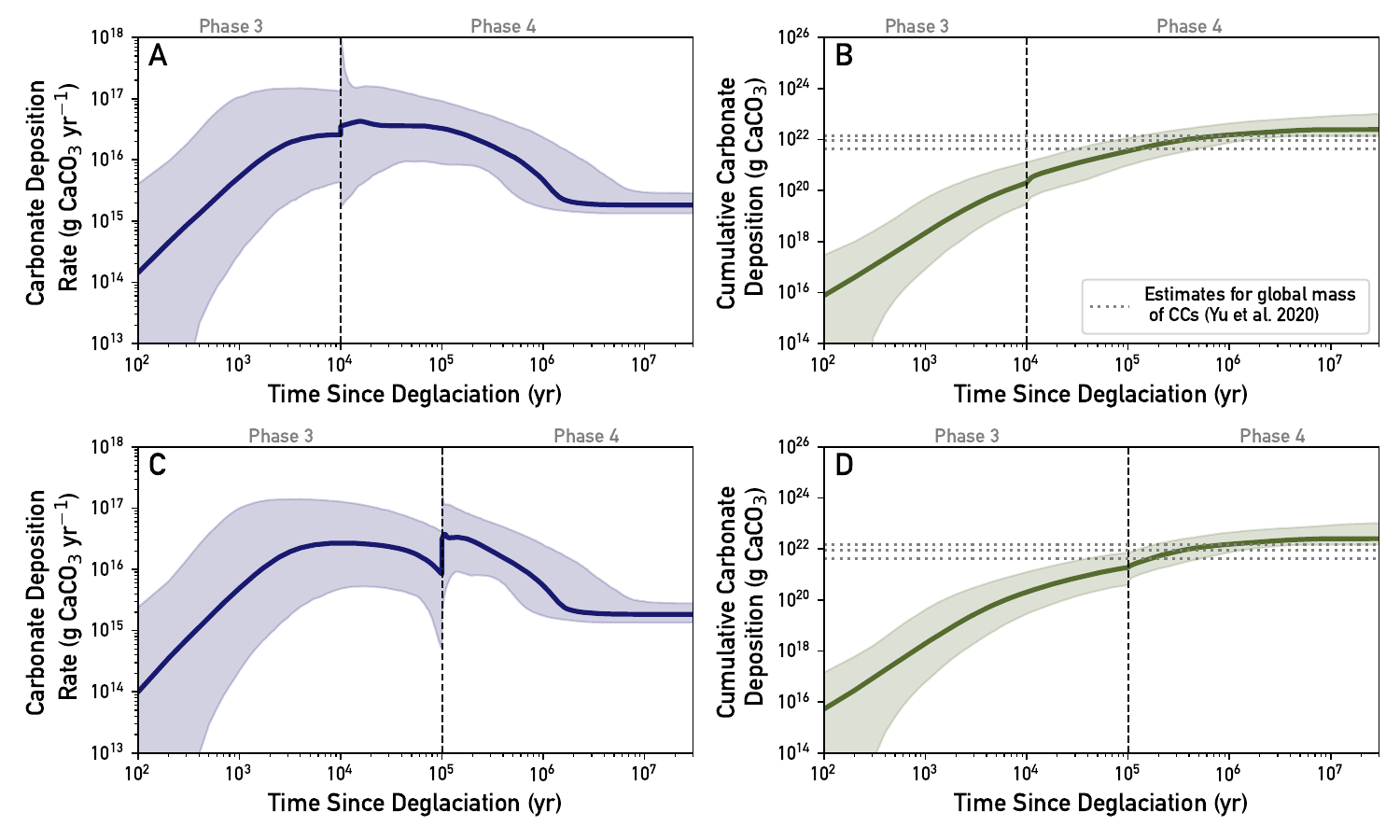}
    \caption{Post-glacial carbonate deposition by mass in the high-alkalinity meltwater scenario. The initial alkalinity of the meltwater in phase 3 is 70,000 $\mu$mol eq kg$^{-1}$, compared to 700 - 1000 $\mu$mol eq kg$^{-1}$ in the baseline case. The end of the stratified ocean (Phase 3), $t_{strat}$, is indicated by the vertical dashed black lines. In panels \textbf{A} and \textbf{B}, $t_{strat} = 10^4$ years. In panels \textbf{C} and \textbf{D}, $t_{strat} = 10^5$ years. Horizontal dotted lines in panels \textbf{B} and \textbf{D} are minimum, median, and maximum estimates of the global mass of Marinoan cap carbonates from \citet{yu_cryogenian_2020}: $\SI{4.2e21}{}$, $\SI{9.3e21}{}$, and $\SI{14.4e21}{g}$, respectively. The shaded regions are the 95\% confidence intervals and the solid lines are the median evolutions with randomly sampled parameters in 3000 model runs.}
    \label{carbdep_mass_highalk}
\end{figure}

Coupled climate and ice-sheet models indicate that the deglaciation process lasted $\sim 2$ kyr, due to the ice-albedo instability \citep[e.g.,][]{hyde_neoproterozoic_2000,myrow_rapid_2018}. We use our silicate weathering parameterization to calculate the delivery of alkalinity from melting glaciers during these 2 kyr. We assume that \ch{pCO2} is 0.1 bar and the surface temperature is 310 K during deglaciation. We fix the weatherability factor, $f_w = 1$, and then vary the empirical factors in their standard ranges. This results in a maximum alkalinity delivery rate of $\SI{4.73e15}{mol.eq.yr^{-1}}$ and a maximum meltwater alkalinity of 70,000 $\mu$mol eq kg$^{-1}$ when integrated over 2 kyr. This is 70-100 times higher than our previous meltwater alkalinity assumption of 700 - 1000 $\mu$mol eq kg$^{-1}$ based on modern glacial meltwater. Moreover, the true value may be even higher because the land surface was likely highly weatherable after enduring millions of years of rock-grinding glacier movement. Here we test this scenario by running our model with the higher initial meltwater alkalinity of 70,000 $\mu$mol eq kg$^{-1}$.

The meltwater layer rapidly saturates and carbonates begin precipitating immediately after deglaciation in this scenario (Figure \ref{carbdep_mass_highalk}). Here, deposition starts 100 years after deglaciation, which is essentially instantaneous in our model, since 100 years is the minimum timestep in this phase. On the other hand, the total timescale of CC deposition is similar to the baseline case. This is because the high-alkalinity meltwater scenario only adds alkalinity to the meltwater layer equivalent to 2 kyr of weathering, which is small relative to the 100 kyr timescale of full CC deposition in the baseline case.

Additionally, the meltwater layer is much less acidic in this scenario (Supplementary Figure S4). In the baseline case, the pH of the meltwater layer starts at $\sim 5$ and then increases to $\sim 7$ after 100 kyr. In the high alkalinity scenario, however, the pH of the meltwater layer starts over 7 and reaches close to 10 in some cases. This is a consequence of inorganic carbon speciation: when all else is held equal, increased alkalinity causes increased pH.

\subsection{Comparison to geologic evidence}

Here we compare our model results to several lines of evidence in the Marinoan CCs, which refers to the combination of (1) the cap dolostones, which are uniform in character, ubiquitous, and sit directly on top of Marinoan glacial deposits, and (2) the overlying limestones, which are more  variable in character and thickness, and sit on top of the dolostones. Our conclusions are therefore most applicable to the Marinoan glaciation. However, some results may generalize to the Sturtian glaciation, which has CCs with notable differences from the Marinoan, such as being often far thinner or absent and generally limestone, not dolostone.

Our results are consistent with the global mass and thickness of the Marinoan CCs. As shown in Figure \ref{carbdep_mass}, we find that enough alkalinity is generated in the post-glacial aftermath to explain the estimated global mass of CCs. The alkalinity source is a combination of continental and seafloor weathering, described above. Considering deposit thickness, we find that the observed global average CC thickness, $\sim11$ meters \citep{yu_cryogenian_2020}, is generally precipitated in under 200 kyr, while thicker sections can be deposited on longer timescales in model runs further from the median evolution (Figure \ref{carbdep_height}). This is consistent with the geologic record because there was likely significant regional variability in depositional environments \citep[e.g.,][]{creveling_sea-level_2014,irie_nonmonotonic_2019} and some CCs are hundreds of meters thick, such as the Noonday Formation \citep[e.g.][]{petterson_neoproterozoic_2011}.

The timescale of CC deposition is constrained by three general types of evidence that have not been reconciled: radiometric dating, paleomagnetism, and sedimentology (Table \ref{timescales}). Although useful for determining absolute ages, the radiometric dating measurements cannot give a statistically significant estimate for the timescale of CC deposition due to the measurement uncertainty and scattered locations in different CCs. The presence of multiple paleomagnetic reversals in multiple sections is evidence for a depositional timescale on the order of $10^5$ to $10^6$ years, with uncertainty due to the unknown frequency of polarity reversals in the Neoproterozoic and the fidelity of the paleomagnetic data and their interpretations. Marinoan CCs also have unusual, heavily debated sedimentary structures. The standard Snowball hypothesis \citep{hoffman_snowball_2017} suggests that features like giant wave ripples and sheet-crack cements are present in CCs as a result of rapid deglaciation, implying depositional timescales of $\sim10^3$ to $10^4$ years \citep[e.g.][]{shields_neoproterozoic_2005, allen_extreme_2005, hoffman_sheet-crack_2010}. However, these features have instead been interpreted as tepees and bedding expansion features that imply prolonged depositional timescales of $\geq10^5$ years \citep[e.g.][]{fairchild.2007,spence_sedimentological_2016,Wallace.2019}.

Our results help reconcile the differing lines of evidence for depositional timescale of the CCs.

First, we predict that the minimum global CC mass is deposited in 32 to 591 kyr in our nominal 95\% confidence intervals. This result suggests that the global timescale of CC deposition was intermediate, with endmember scenarios for extremely rapid or prolonged deposition due to regional variability corresponding to model runs outside of the 95\% confidence interval. So, we suggest that the sedimentary structures may reflect deposition on both short and long timescales. On one hand, deposition on timescales below 30 kyr may have occurred in regions with enhanced alkalinity delivery, where deposition can start as fast as 100 years after deglaciation (Figure \ref{carbdep_mass_highalk}). On the other hand, deposition on timescales over 500 kyr can be reconciled by the fact that our modeled deposition rates remain higher than the baseline for over 2 Myr following deglaciation (Figure \ref{carbdep_mass}). Considering the regional variability, it is likely that the overlying limestones in several sections experienced prolonged deposition in favorable environments - i.e., the Brazil, Oman, and Australian CCs - thus explaining the presence of the paleomagnetic reversals and some of the sedimentary structures.

Second, we predict that the peak carbonate deposition rate is achieved on timescales of $10^4$ yrs after deglaciation (Figures \ref{carbdep_mass} and \ref{carbdep_mass_highalk}). We suggest that this can help reconcile evidence of both rapid and prolonged deposition. In our interpretation, the $10^3$-$10^4$ yr timescale of CC deposition inferred from some of the sedimentary structures is an underestimate because it is assumed that the peak deposition rates lasted for the entirety of CC deposition. Perhaps some of the sedimentary structures indeed indicate peak deposition rates, but they did not last for the entirety of CC deposition, creating other sedimentary structures and recording paleomagnetic reversals as a result of longer deposition. This idea is consistent with our model results and a recent analysis of the Svalbard Marinoan CC \citep{Fairchild.2023}, which both indicate peak deposition rates occurring shortly after deglacation and then declining over time. The record of peak deposition rates can be further analyzed by comparing the distribution of sedimentary structures with respect to height to our qualitative deposition rates as a function of height (Figure \ref{carbdep_height}).

Third, our results are consistent with the contact between glacial deposits and CCs. In all Marinoan sections there is sharp contact between CC and glacial unit \citep[e.g.,][]{hoffman_snowball_2017}, suggesting that the CCs precipitated during and immediately after deglaciation, with no hiatus \citep[e.g.][]{Domack.2011}. In the high-alkalinity meltwater scenario (Figure \ref{carbdep_mass_highalk}), we show that there is no depositional hiatus (i.e. no more than $\sim100$ years). In the baseline scenario with low initial meltwater alkalinity (which is biased toward slow deposition relative to the high-alkalinity meltwater scenario), the hiatus only lasts $\sim1$-$\sim10$ kyr, which may be consistent with the evidence depending on sedimentation rates. Our model does not explicitly treat the dynamics of the deglaciation with respect to changing meltwater volume, regional variability, and weathering rates, so the two model scenarios cannot distinguish between syn-deglacial or post-deglacial deposition, but they are consistent with the general lack of a depositional hiatus, which has been a longstanding problem for CC explanations involving post-glacial continental weathering as an alkalinity source \citep[see][]{yu_cryogenian_2020}.

Our results are also consistent with the general stratigraphy of the Marinoan CCs. Our model shows that CC deposition in the meltwater layer (Phase 3) is caused by alkalinity supply during and immediately after deglaciation, which is consistent with the transgressive nature of the Marinoan cap dolostones, deposited as sea levels rose \citep{hoffman_snowball_2002}. Subsequent CC deposition in the well-mixed ocean (Phase 4) is consistent with the more variable and prolonged deposition of the overlying limestones, as outlined in the ``cap limestone'' phase of \citet{shields_neoproterozoic_2005}; here we have shown it is valid in a global geochemical model.

\begin{figure}[htbp]
    \centering
    \includegraphics[width=\textwidth]{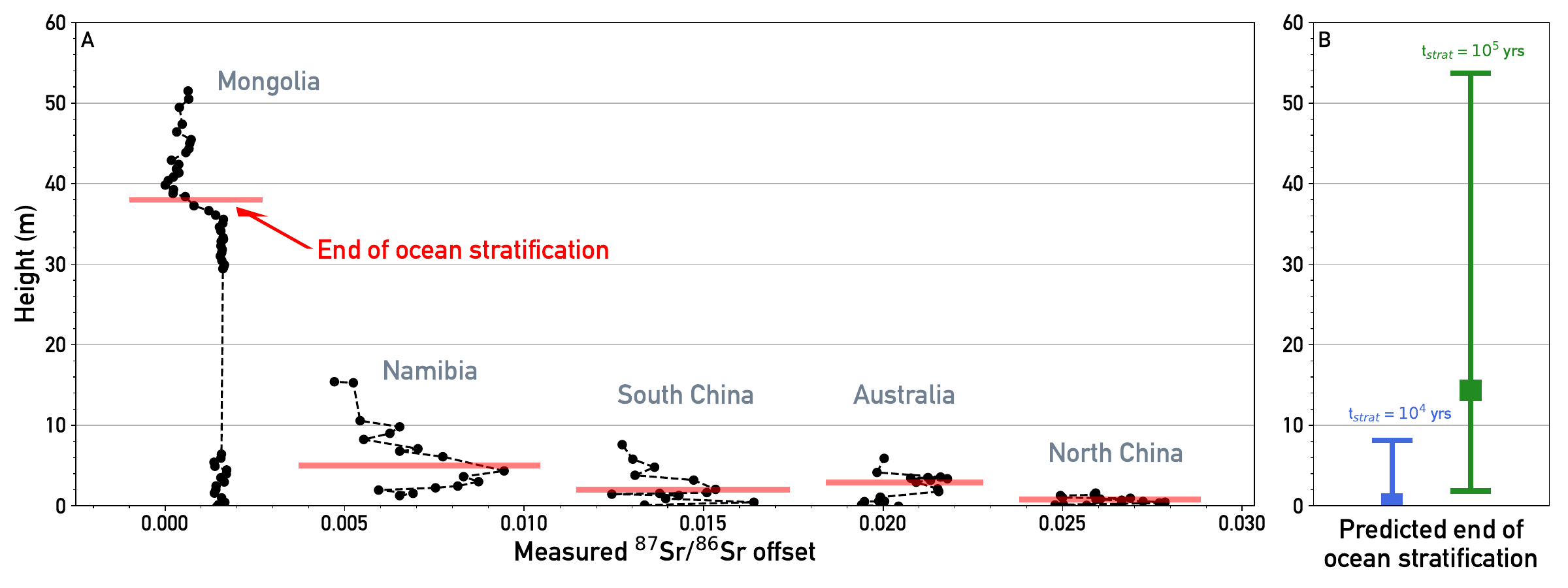}
    \caption{Comparison of measured strontium isotopes and model results for determining the end of post-glacial ocean stratification. \textbf{A}, Measured $^{87}$Sr/$^{86}$Sr offsets upsection in 5 cap carbonates. The red lines are the roughly inferred heights in the cap carbonate at which ocean stratification ends according to the mixing  models of those who obtained and analyzed the data: \citet{liu_neoproterozoic_2014, liu_sr_2018} (Mongolia), \citet{liu_geochemical_2013} (Australia), and \citet{wei_ca_2019} (Namibia, South China, and North China). Below the red lines, the cap carbonates were precipitated out of mostly glacial meltwater with elevated, continentally influenced $^{87}$Sr/$^{86}$Sr. Above these lines, the cap carbonates were precipitated out of mostly ocean water with lower, hydrothermally influenced $^{87}$Sr/$^{86}$Sr.  \textbf{B}, Model predictions for the height at which ocean stratification ends in the baseline scenario. These are effectively predictions for where the red lines should be. Results assuming $t_{strat} = 10^4$ years are on the left in blue, and assuming $t_{strat} = 10^5$ years are on the right in green. The filled squares are the median model predictions and the error bars are the 95\% confidence interval corresponding to Figure \ref{carbdep_height}.}
    \label{strontium}
\end{figure}

Several lines of geochemical evidence probe Marinoan post-glacial ocean stratification. The $^{87}$Sr/$^{86}$Sr ratio in seawater is raised by continental weathering and lowered by hydrothermal input \citep[e.g.,][]{davis_imbalance_2003,vance_variable_2009,allegre_fundamental_2010}. Several studies \citep{liu_sea_2013,liu_neoproterozoic_2014,liu_sr_2018,wei_ca_2019} have measured $^{87}$Sr/$^{86}$Sr along CC sections and found that they show a stepwise decrease from high to low values upsection, indicated by the red lines in Figure \ref{strontium}a. This decrease suggests that (1) first, CCs precipitate out of a glacial meltwater layer with elevated $^{87}$Sr/$^{86}$Sr due to massive continental weathering, and then (2) second, the CCs precipitate out of a well-mixed ocean which has a lower $^{87}$Sr/$^{86}$Sr that is closer to the typical modern ocean value. Thus, the height at which this stepwise decrease occurs should reflect the point at which the post-glacial ocean becomes well-mixed. This interpretation is further supported by other Ca, Mg, and Sr isotope measurements \citep{wei_ca_2019,ahm_early_2019,wang_application_2023}.

Our results are consistent with $^{87}$Sr/$^{86}$Sr trends in CC sections. In Figure \ref{strontium}b, we show our model results for the CC height at which ocean stratification ends. In the case with $t_{strat} = 10^4$ years, 0.18 meters of the CCs are deposited in the stratified ocean in the median model run; with $t_{strat} = 10^5$ years, this value increases to 14.36 meters. The 95\% confidence intervals of these two cases incorporate the full range of Sr-inferred mixing heights. Our model is global, so it is likely that different places experienced different timescales of ocean stratification. For example, we would expect that the Mongolian deposit experienced $10^5$ years or more of ocean stratification, the North China deposit experienced $10^4$ years of ocean stratification or even less, and the Namibia, South China, and Australia deposits experienced an intermediate time between $10^4$ and $10^5$ years.

\section{Discussion} \label{disc}

\begin{figure}[htbp]
    \centering
    \includegraphics[width=0.6\textwidth]{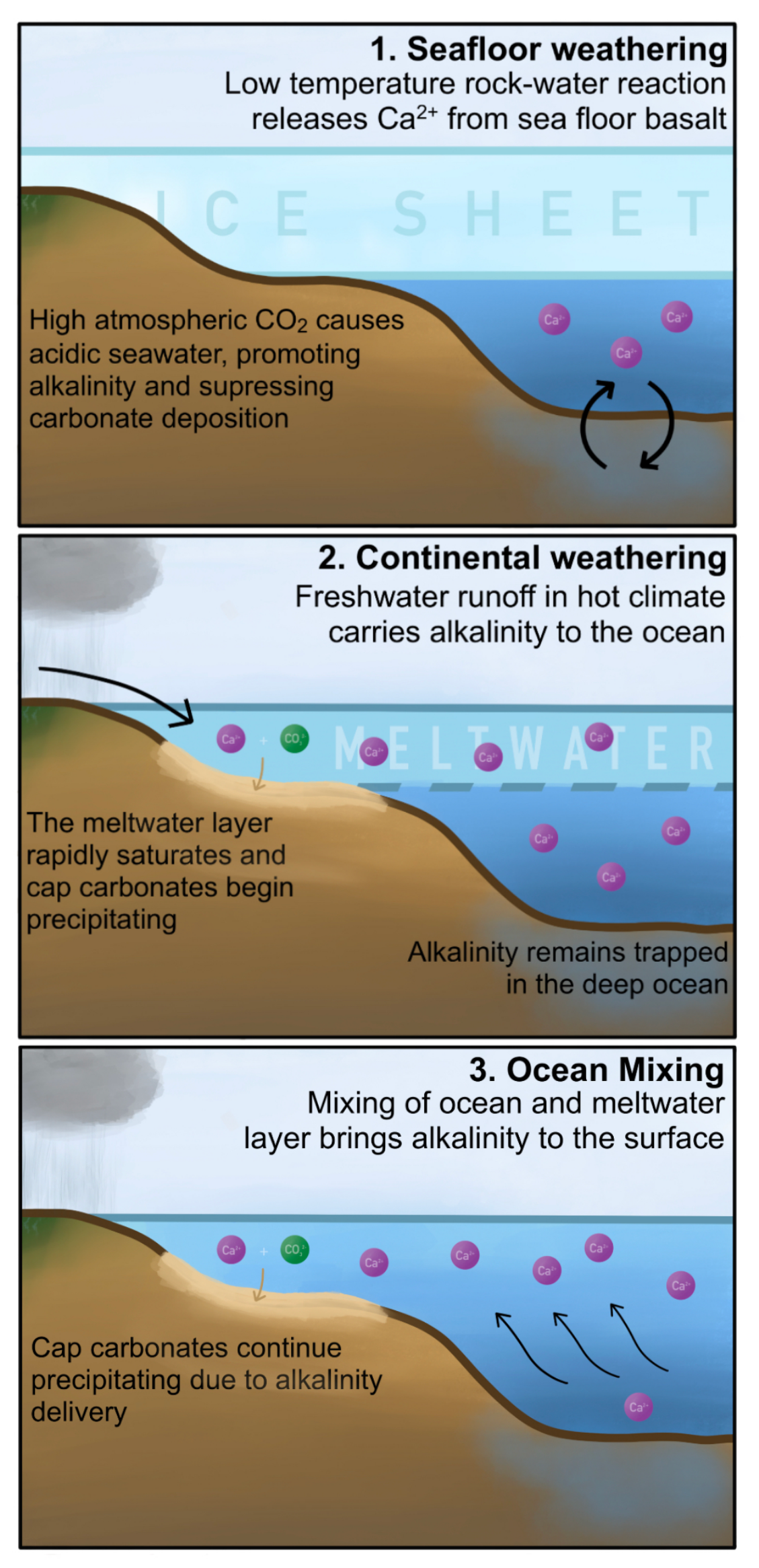}
    \caption{Schematic diagram of the ``\textbf{S}eafloor weathering-\textbf{C}ontinental weathering-\textbf{O}cean \textbf{M}ixing (SCOM)'' mechanism for cap carbonate deposition.}
    \label{scom}
\end{figure}

In the above sections we have presented a global model for the deposition of CCs after the Marinoan glaciation. This model is consistent with geologic evidence regarding the global mass and thickness of the CCs, the timescale of their deposition, and their deposition in a post-glacial stratified ocean. Our results could be summarized as the ``Seafloor weathering-Continental weathering-Ocean Mixing (SCOM)'' mechanism for CC deposition (Figure \ref{scom}).

The SCOM mechanism has 3 phases. First, seafloor weathering in high \ch{CO2}, acidic conditions during the glaciation supplies alkalinity to the sub-glacial ocean. Second, intense continental weathering supplies alkalinity to the post-glacial meltwater layer and CCs begin precipitating. Third, the eventual mixing of the meltwater layer and the deep ocean supplies alkalinity to the surface and CCs continue precipitating.

The SCOM mechanism explains Marinoan CC deposition on a global scale and was likely subject to regional variability. For example, post-glacial sea level rise was likely highly regionally variable \citep[e.g.,][]{creveling_sea-level_2014,irie_nonmonotonic_2019}, runoff rates from continental weathering depend on regional topography, and the duration of ocean stratification likely depends on regionally varying currents. These important regional distinctions explain how individual CCs could deviate from our median model predictions, but even the most unique CCs are broadly consistent with our 95\% confidence intervals. We therefore propose that Marinoan CC deposition on a global scale was primarily driven by the SCOM mechanism. 

The SCOM mechanism draws from several previous hypotheses for CC deposition. In the plumeworld hypothesis \citep{shields_neoproterozoic_2005}, it was proposed that CCs rapidly precipitate out of a stable glacial meltwater layer with alkalinity generated from a variety of sources. Many others have hypothesized that rapid continental weathering would have supplied the necessary alkalinity to the ocean to generate the CCs \citep[e.g.,][]{hoffman_snowball_2002,higgins_aftermath_2003}. Deep ocean upwelling of alkalinity - derived from the degradation of organic matter, not seafloor weathering - has also been suggested \citep{grotzinger_anomalous_1995,knoll_comparative_1996}. While these hypotheses are plausible and have been extensively studied individually, they struggle to explain the timescale of deposition, the required alkalinity source, and other geologic evidence \citep{yu_cryogenian_2020}. With the SCOM mechanism, we self-consistently combine and refine these hypotheses to show that it is consistent with the geologic evidence on a global scale. The key improvements presented here include the previously unconsidered alkalinity source of seafloor weathering, the explicit calculation of aqueous chemistry in the glacial meltwater layer and subsequent mixed ocean, and the rigorous calculation of the geologic carbon cycle.

There are several caveats and limitations on the SCOM mechanism from both the geologic record and our modeling approach. First, transitions between phases are imposed in our model due to their complexity; for example, the deglacial transition would require a careful treatment of ice sheet dynamics and climate, and the ocean mixing transition would require GCM-like treatment of 3D ocean dynamics. We do, however, provide a sensitivity study to explore the effect of a prolonged deglaciation (Supplementary Section C2). In the most extreme endmember, this may increase the time required for CC deposition by up to 90 kyrs, but does not change our main conclusions. Further discussion of forcing model transitions and sensitivities are found in Supplementary Section C. Second, our model is global, and does not explicitly capture regional behavior, which is important for interpretation of individual CCs. Third, compilation of the global CC record is subject to variability in interpretation and measurement, which may cause biases (e.g., toward thicker CC sections than are actually present or toward overinterpetation of paleomagnetic data). Future studies on the model transitions, the implications for regional behavior, and improved global interpretation of the CC record can improve or test the SCOM mechanism and are encouraged.

There are many other lines of geologic evidence that can be used to test the SCOM mechanism, including  ocean pH proxies of boron isotopes \citep[e.g.][]{Kasemann.2010} and rare Earth element distributions \citep{Lin.2024}, styles of carbonate deposition within CCs, sedimentation rates and carbonate concentrations as a function of height within CCs, developing proxies that probe alkalinity in the subglacial ocean, and more radiosotope dates from multiple CCs (preferably at the top and bottom of the same section). Additionally, comparisons to the record of Sturtian CCs may provide insights into how the two major Cryogenian glaciations differed (e.g., the effect of disparate glacial durations) and their implications on the evolution of Earth's surface. In this work we have shown that the SCOM mechanism is broadly consistent with the global characteristics of Marinoan CCs, but future studies making more detailed, regional comparisons to the above evidence are suggested.

\section{Methods}


We model the evolution of the carbon content and marine alkalinity in Earth's atmosphere and ocean as it is subject to processes of the geologic carbon cycle before, during, and after a global glaciation event. Our model is based on previous geologic carbon cycle models that have been rigorously validated against Earth's last 100 Myr and applied as far back as 4 Ga \citep{krissansen-totton_constraining_2017,krissansen-totton_constraining_2018}. In order to capture the unique climate of the Cryogenian glaciations, we modify the model accordingly and separate it into 4 distinct configurations (Figure \ref{model_schematic}): the background Neoproterozoic phase, a syn-glacial phase, a post-glacial stratified ocean phase, and a post-glacial well-mixed ocean phase.

Key aspects of the model are described below. Additional information is in the Supplementary Material. The calculation of the baseline carbon cycle fluxes, aqueous chemistry, and climate is in Supplementary Section A. Calibration and benchmarking of the model against the modern and post-glacial Earth is in Supplementary Section B. Further description of transitions between model phases and a sensitivity study are in Supplementary Section C.

\subsection{Phase 1: Background Neoproterozoic}

The goal of phase 1 is to find the background climate in the Neoproterozoic era before any glaciation events occur. In this phase, there are two model reservoirs: one for the combined atmosphere and ocean, and one for water in the pore-space of the seafloor. The pore-space reservoir captures the chemistry of water circulating through the the upper portion of the oceanic crust, which is important for investigating the impact of seafloor weathering. The time evolution of carbon chemistry in these reservoirs is described by the following set of equations:
\begin{equation} \label{neosystem}
\begin{split}
\frac{dC_o}{dt} & = \frac{1}{M_o}\Bigl( -J({\rm DIC_o} - C_p) + V_{\rm total} + W_{\rm carb} - P_{\rm shelf} \Bigr) \\
\frac{dA_o}{dt} & = \frac{1}{M_o}\Bigl( -J(A_o - A_p) + 2W_{\rm sil} + 2W_{\rm carb} - 2P_{\rm shelf} \Bigr) \\
\frac{dC_p}{dt} & = \frac{1}{M_p}\Bigl( -J(C_p - {\rm DIC_o}) - P_{\rm pore} \Bigr) \\
\frac{dA_p}{dt} & = \frac{1}{M_p}\Bigl( -J(A_p - A_o) + 2W_{\rm sea} - 2P_{\rm pore} \Bigr). \\
\end{split}
\end{equation}

Here, $C$ is the concentration of inorganic carbon with units mol C kg$^{-1}$ and $A$ is the carbonate alkalinity with units mol eq kg$^{-1}$ where the subscript $o$ and $p$ indicate the ocean-atmosphere and pore-space reservoirs, respectively. $C_p$ is equal to the dissolved inorganic carbon (DIC) of the pore-space. $C_o$ is the sum of DIC in the ocean and the carbon content in the atmosphere, given by $C_o = \ch{DIC}_o + \ch{pCO2} \times s$, where $s$ is a scaling factor equal to the total number of moles C per bar in the atmosphere divided by the mass of the ocean, $s = \num{1.8e20}/M_o$, and \ch{pCO2} is in bar. We assume that the masses of the ocean and pore-space water are equal to their modern values, respectively given by $M_o = \num{1.35e21}$ kg and $M_p = \num{1.35e19}$ kg \citep{caldeira_long-term_1995}. $J$ is the water mass flux between the deep ocean and pore-space, which has been estimated by balancing crustal heat fluxes: $J = 0.6-\SI{2e16}{kg.yr^{-1}}$ \citep{coogan_low-temperature_2018}; this is equivalent to the entire ocean circulating through the pore space every 70 to 250 kyrs.

The remaining terms are fluxes of carbon (mol C yr$^{-1}$) and alkalinity (mol eq yr$^{-1}$) due to processes of the geologic carbon cycle: $V_{\rm total}$ is the sum of volcanic outgassing from subaerial and mid-ocean ridge sources, $W_{\rm carb}$ is continental carbonate weathering, $W_{\rm sil}$ is continental silicate weathering, $W_{\rm sea}$ is seafloor weathering, $P_{\rm shelf}$ is carbonate precipitation on the continental shelf, and $P_{\rm pore}$ is carbonate precipitation in the pore-space. Calculation of these fluxes is in Supplementary Section A.

\subsection{Phase 2: Syn-glacial}

Starting from the background Neoproterozoic climate, we impose a glaciation event. In the syn-glacial phase, we calculate the time-dependent evolution of the geologic carbon cycle until a threshold for deglaciation is reached. The equations in \ref{neosystem} are still the governing equations of this phase, but they are modified in several ways, described below.

With the presence of a large global ice sheet, sea level should fall significantly and the global mass of the liquid ocean should decrease. \citet{hoffman_snowball_2017} estimate ocean volume change during glaciation by summing continental and sea ice volumes under various dust accumulation rates and \ch{pCO2} levels \citep{goodman_feedbacks_2013, benn_orbitally_2015}. We explore the range of estimates and assume a decrease in ocean volume of 10\% to 30\% relative to the modern ocean. To calculate the immediate increase in dissolved species concentration from a shrinking ocean, we assume that no conservative cations (e.g., \ch{Ca^{2+}}) are trapped in the ice during freezing. Thus, the ocean alkalinity and dissolved inorganic carbon increase relative to the background Neoproterozoic values by the same relative proportion that the ocean volume decreases.

The 2D and 3D ocean circulation models of \citet{ashkenazy_dynamics_2013, ashkenazy_ocean_2014} show that the sub-glacial ocean should have had vigorous convective mixing under the ice cover, causing it to be isothermal and chemically well-mixed. Thus, we continue to use a single box in our model to represent the ocean. We assume that the syn-glacial ocean had a uniform and constant temperature of 269.5 K, regardless of \ch{pCO2}, which is indicated by ocean models with complete ice cover \citep{ashkenazy_dynamics_2013, ashkenazy_ocean_2014}. 
This assumption is justified for scenarios with incomplete ice cover as well since global averaged surface temperatures are predicted to be below 269.5 K \citep{hoffman_snowball_2017}.

We nominally assume that the atmosphere and sub-glacial ocean were in equilibrium with respect to inorganic carbon speciation and aqueous chemistry during the glaciation. Several lines of evidence support this assumption: (1) It has been shown that atmosphere-ocean equilibrium with respect to \ch{CO2} can be reached on million year timescales with only $10^3$ km$^2$ open ocean \citep{le_hir_scenario_2008}. Areas of open ocean likely exceeded this threshold via geothermal heat production and lava flows alone, as the modern area of emerged active hydrothermal systems ($\SI{1.5e6}{km^2}$) is 3 orders of magnitude greater than what is required for equilibration \citep{dessert_basalt_2003}. (2) Sedimentological evidence, oxygen isotopes, and sulphur isotopes of the Svalbard Marinoan glacial deposits indicate that ice sheets were sensitive to orbital forcing as \ch{pCO2} rose, creating extensive patches of open water \citep{mitchell_orbital_2021}. (3) Geochemical measurements of carbon, nitrogen, and iron in the Nantuo Formation suggest there was aerobic nitrogen cycling in surface waters and swaths of open ocean at mid-latitudes during glaciation \citep{song_mid-latitudinal_2023}. (4) Approximately 25\% of global volcanic outgassing of \ch{CO2} occurs at underwater mid-ocean ridges on the modern Earth \citep[][p. 203]{catling_atmospheric_2017}. \ch{CO2} bubbles from this volcanism would directly equilibrate with the subglacial ocean regardless of ocean-atmosphere equilibrium. Combined, this evidence indicates that a chemically isolated atmosphere and ocean is unlikely. 

In order to determine the minimum timestep for modeling this phase, we must estimate how long it takes for the atmosphere-ocean equilibrium to be reached. \citet{le_hir_scenario_2008} showed that only 3000 km$^3$ of open ocean is required for full ocean \ch{CO2} diffusion on the order of several millions of years. When combined with the efficient mixing of the subglacial ocean, it is reasonable to assume that equilibrium would be reached on timescales similar to the modern ocean. We conservatively assume equilibration takes 100 times longer than the modern time, and thus our minimum model timestep during this phase is 100,000 years.

During widespread ice sheet coverage on land, continental weathering rates are expected to have been significantly reduced or completely stopped. We nominally assume that there was no continental weathering during glaciation; however, we do test our model with the incorporation of low, constant weathering rates calculated in \citet{lan_massive_2022}. Their models estimate that the continental silicate weathering rate was $3.3-22 \times 10^8$ mol eq yr$^{-1}$ and that the continental carbonate weathering rate was $8-18 \times 10^8$ mol C yr$^{-1}$, both of which are several orders of magnitude lower than their modern rates of $\sim\SI{8e12}{mol.eq.yr^{-1}}$ \citep{moon_new_2014} and $\sim\SI{11e12}{mol.C.yr^{-1}}$ \citep{hartmann_global_2009}, respectively. Our parameterization of seafloor weathering and pore-space carbonate deposition remains unchanged through this time period.

There should be few neritic environments for carbonate deposition in the syn-glacial ocean. Sea level is expected to have dropped by over 500 m during glaciation due to the size and extent of the ice sheets \citep{liu_sea_2013}, so the syn-glacial sea level should be too low to support carbonate deposition on continental shelves given that they are generally not deeper than 200 m on modern Earth. However, evidence has been found for syn-glacial carbonate deposition during the Sturtian glaciation \citep{hood_neoproterozoic_2022}, indicating that some carbonate depositional environments must have persisted through global glaciation, potentially on the continental slopes. This motivates us to allow carbonate deposition during glaciation. We continue to use the baseline parameterization to calculate the deposition rate, but we reduce the ratio of syn-glacial continental shelf area to modern ($A_{\rm shelf}/A^{\rm mod}_{\rm shelf}$) to account for the decrease in depositional environments. On the modern Earth, the surface area of continental shelves is $\SI{3.22e7}{km^2}$ and of continental slopes is $\SI{1.96e7}{km^2}$ \citep{harris_geomorphology_2014}; the removal of continental shelves results in a 62\% decrease in depositional area. Thus, we explore a range of $A_{\rm shelf}/A^{\rm mod}_{\rm shelf}$ values from 0.3 to 0.5 to simulate the elimination of deposition on continental shelf environments during glaciation.

\subsection{Phase 3: Post-glacial Stratified Ocean}

After the syn-glacial phase ends, the model transitions to the post-glacial stratified ocean phase. This phase is time-dependent and spans from the end of the glaciation until the ocean becomes well-mixed.

To account for the post-glacial ocean stratification, we alter the Neoproterozoic background model for atmosphere and ocean chemistry by splitting the whole-ocean box into two boxes: one box contains the atmosphere and meltwater layer, and the other contains the deep ocean. We follow \citet{boudreau_secular_2019} to design the basic framework of the 2-box model. The time evolution of carbon chemistry is now described by the following set of equations:

\begin{equation} \label{postsystem}
    \begin{split}
    \frac{dC_s}{dt} & = \frac{1}{M_s}\Bigl( -K({\rm DIC_s} - C_d) + V_{\rm air} + W_{\rm carb} - P_{\rm shelf} - O_{\rm sink} \Bigr) \\
    \frac{dA_s}{dt} & = \frac{1}{M_s}\Bigl( -K(A_s - A_d) + 2W_{\rm sil} + 2W_{\rm carb} - 2P_{\rm shelf} \Bigr) \\
    \frac{dC_d}{dt} & = \frac{1}{M_d}\Bigl( -K(C_d - {\rm DIC_s}) - J(C_d - C_p) + V_{\rm ridge} + O_{\rm sink} \Bigr) \\
    \frac{dA_d}{dt} & = \frac{1}{M_d}\Bigl( -K(A_d - A_s) - J(A_d - A_p) \Bigr) \\
    \frac{dC_p}{dt} & = \frac{1}{M_p}\Bigl( -J(C_p - C_d) - P_{\rm pore} \Bigr) \\
    \frac{dA_p}{dt} & = \frac{1}{M_p}\Bigl( -J(A_p - A_d) + 2W_{\rm sea} - 2P_{\rm pore} \Bigr). \\
    \end{split}
\end{equation}

The model now tracks carbon and alkalinity in three boxes: the surface ocean meltwater layer (s), the deep ocean (d), and the pore-space (p). $C$ is still the concentration of inorganic carbon and $A$ is still carbonate alkalinity. $C_s$ contains the sum of the DIC in the surface ocean and the carbon in the atmosphere, $C_s = \ch{DIC}_s + \ch{pCO2} \times s$, and $s$ is adjusted to reflect the mass of the surface ocean meltwater layer, $s = \num{1.8e20}/M_s$. $J$ is the water mass flux between the deep ocean and pore-space, and is unchanged from the Neoproterozoic background configuration. $K$ is the water mass flux between the surface and deep ocean. The volcanic outgassing flux is now split into a subaerial component ($V_{\rm air}$) and mid-ocean ridge component ($V_{\rm air}$), which are 75\% and 25\% of the total volcanic flux, respectively \citep[][p. 203]{catling_atmospheric_2017}. Finally, $O_{\rm sink}$ is the flux of sinking organic carbon from the surface to the deep ocean.

We assume that the deep ocean retains the same DIC and alkalinity as the final results from the sub-glacial ocean in phase 2. On the other hand, we must make assumptions about the initial chemical properties of the glacial meltwater. To determine the alkalinity, we use compositional measurements of modern glacial meltwater, compiled in \citet{brown_glacier_2002}. The compilation includes 22 different measurements of glacial meltwater from around the world, including both land and sea glaciers. The average alkalinity from the minimum and maximum bounds of these measurements is 700-1000 $\mu$mol eq kg$^{-1}$. We take this as the starting alkalinity for the post-glacial surface ocean. We then assume that the meltwater is immediately equilibrated with the atmosphere to calculate the DIC.

We assume the total post-glacial ocean mass is equal to the modern day ocean mass, as we did in phase 1. This assumption is justified because modern Earth's glaciers are responsible for only $\sim 2\%$ of the total water budget \citep{pilson_introduction_2012}, so even if they were completely melted in the post-glacial hothouse climate they would not significantly change the mass of the ocean. The mass of the meltwater layer is 10\%-30\% of the global ocean mass, consistent with the ice volume in phase 2. The deep ocean makes up the rest of the total mass.

We assume the glacial meltwater layer remains chemically distinct from the deep ocean for a variable amount of time after deglaciation. We assume the stratification lasts for $t_{strat} = 10^4$ or $10^5$ years, consistent with 1D and 3D models of the post-glacial ocean \citep{yang_persistence_2017,ramme_climate_2022}. These models form the basis for the assumptions we make to simulate the post-glacial ocean. First, we nominally assume that there is no mixing between the meltwater layer and the deep ocean ($K=0$) during stratification. Second, we assume that the meltwater layer is in thermal equilibrium with the surface, as the melting glacial water is quickly heated at the surface before sinking deeper into the layer. Third, we assume that the deep ocean temperature warms linearly during the time of stratification. As is shown in the models, the ocean eventually recovers to a well-mixed state with a surface temperature for which our baseline deep ocean temperature parameterization applies. Thus, the deep ocean temperature ($T_d$) evolution is described by
\begin{equation}
    T_d = 269.5  + \frac{t}{t_{strat}} \Bigl( a_{\rm grad}T_s + b_{\rm int} - 269.5 \Bigr)
    \label{substrattemp}
\end{equation}
where 269.5 K is the temperature of the sub-glacial ocean, $t$ is the time after glaciation, and the term in the parenthesis is the difference between the deep ocean temperature calculated via the baseline parameterization and the sub-glacial ocean temperature.

The minimum timestep for this phase is set by the time it takes for chemical equilibrium to be reached between the meltwater layer and the atmosphere. We expect this to be shorter than the modern timescale for several reasons. First, the meltwater originates at the surface in small parcels as it melts. We expect that the meltwater should reach atmospheric equilibrium as soon as it is produced due to the high surface area to volume ratio of the parcels and their direct exposure to the atmosphere. Second, the meltwater layer contains only the upper portion of the ocean. Mixing at the surface is more vigorous than deep ocean mixing due to e.g. wind-driven perturbation and should lead to faster equilibration. This mixing is enhanced when paired with an estimated 22\% increase in runoff relative to modern \citep{le_hir_snowball_2009}. Third, and most importantly, there is simply less water to equilibrate. The meltwater layer has a water mass that is 10-30\% of the modern ocean. Combining these arguments, we indeed assume that equilibration in the meltwater layer is faster than in the modern ocean. We nominally assume this equilibration happens in 100 years, which is the minimum allowable timestep for phase 3.

We apply our baseline continental weathering parameterizations to the post-glacial environment. The post-glacial environment likely had a high \ch{CO2} atmosphere, leading to high surface temperatures and a rapid hydrologic cycle. \citet{le_hir_snowball_2009} investigated the post-glacial climate using the FOAM General Circulation Model and the WITCH weathering model to estimate continental weathering rates as a function of \ch{pCO2}. As shown in Supplementary Section B, our standard continental weathering parameterization aligns with and encompasses the rates from the more detailed models in \citet{le_hir_snowball_2009}. The spread in our derived rates is mainly a product of the weatherability factor, $f_w$, which was varied from 0.5 to 1.5. Narrowing this range can more closely align our predicted rates with those from \citet{le_hir_snowball_2009} but may not be justified given all the uncertain factors that determine $f_w$.

We apply the baseline carbonate deposition parameterization, which was originally developed for the Neoproterozoic pre- and post-glacial climates \citep{ridgwell_carbonate_2003}. The post-glacial shelf area relative to modern is needed in order to apply this parameterization. In the post-glacial climate, the eustatic sea level should be higher in general due to total ice melting and thermal expansion \citep[e.g.,][]{ramme_climate_2022}, but regional-scale sea level change is highly variable \citep{creveling_sea-level_2014}. If all of the ice on modern Earth was melted, it would result in at least 60 m of global sea level rise \citep{frederikse_causes_2020}. Sea level rise due to thermal expansion in the post-glacial hothouse climate is predicted to be up to 8 m \citep{ramme_climate_2022}. Based on modern Earth's hypsometry \citep{amante_etopo1_2009}, a 70 m rise in sea level would cover about $\SI{1.5e7}{km^2}$ of land, which is equivalent to $\sim50\%$ of the area of modern continental shelves \citep{harris_geomorphology_2014}. Using this as the basis for our calculation, post-glacial shelf area would have been 1.5 times larger than the modern area. Thus, we explore the range of values $A_{\rm shelf}/A^{\rm mod}_{\rm shelf} = 1-2$.

We nominally assume that all carbonate precipitation chemistry is based on calcite and aragonite. Post- and syn-glacial snowball Earth carbonates are frequently dolomitic in composition; however, recent isotopic evidence suggests the cap carbonates were deposited as calcium carbonate and later altered to dolomite via marine diagenesis \citep{ahm_early_2019}. Furthermore, a similar modeling study \citep{hood_neoproterozoic_2022} shows that the geochemical evolution of the ocean during and after the Sturtian glaciation is not sensitive to the difference in carbonate chemistries based on using calcite, magnesite, or dolomite as the representative carbonate mineral. Thus, we believe our assumption is justified for modeling the large scale features of the geologic record such as the total mass and depositional timescale of the Cryogenian carbonates. 

Because we have split the ocean box into surface and deep components, we now include an organic carbon export flux from the surface ocean to the deep ocean, following \citet{boudreau_secular_2019}. The modern global rate of organic carbon export is on the order of 100 Tmol yr$^{-1}$ \citep{pilson_introduction_2012}. We expect carbon export in the post-glacial ocean to be greatly diminished for two reasons: (1) total biomass in the Precambrian era should have been less than in the modern day, and (2) the abrubt swings of temperature, ocean pH, and unfrozen ocean area should have exterminated a large portion of existing biomass. Nevertheless, we explore a range of values for carbon export in the post-glacial ocean: $O_{\rm sink} = 50 - 200$ Tmol yr$^{-1}$. We assume that all carbon exported from the surface is regenerated in the deep ocean.

\subsection{Phase 4: Post-glacial well-mixed ocean}

After the post-glacial stratified ocean phase is terminated, the model transitions to the post-glacial well-mixed ocean phase. This phase is time-dependent and spans from the end of the stratified ocean until the background climate state is recovered. The model configuration in phase 4 is almost identical to phase 1. The only difference is the increase in assumed shelf area due to sea level rise, described in phase 3. Phase 4 is terminated when a steady state background climate is reached and the model run is complete.

\section{Data Availability}

The datasets corresponding to this study and code to reproduce all figures are available at Zenodo (doi.org/10.5281/zenodo.12786460). 

\section{Code Availability}

The GOOSE model code developed and used in this work is persistently available at Zenodo (doi.org/10.5281/zenodo.12786460) and on the lead author's GitHub (github.com/trentagon).

\section*{Acknowledgements}

T.B.T. acknowledges funding from the NSF GRFP (DGE-1762114). D.C.C. acknowledges support from NASA Exobiology Program grant no. 80NSSC21K0476. This work is supported in part by the Virtual Planetary Laboratory, a member of NASA NExSS, funded via the NASA Astrobiology Program (Grant 80NSSC18K0829).

\section*{Author Contributions Statements}

T.B.T and D.C.C. designed the project and wrote the manuscript. T.B.T. performed the modeling and analysis.

\section*{Competing Interests Statement}

The authors declare no competing interests.

\end{document}


\title{Supplementary Material for

``Three-stage Formation of Cap Carbonates after Marinoan Snowball Glaciation Consistent with Depositional Timescales and Geochemistry''}

\author[0000-0003-2457-2890]{Trent B. Thomas}
\affiliation{Department of Earth and Space Sciences, University of Washington, Seattle, WA, USA}
\affiliation{Astrobiology Program, University of Washington, Seattle, WA, USA}

\author[0000-0001-5646-120X]{David C. Catling}
\affiliation{Department of Earth and Space Sciences, University of Washington, Seattle, WA, USA}
\affiliation{Astrobiology Program, University of Washington, Seattle, WA, USA}

\section*{}

\begin{figure}[htbp]
    \centering
    \includegraphics[width=\textwidth]{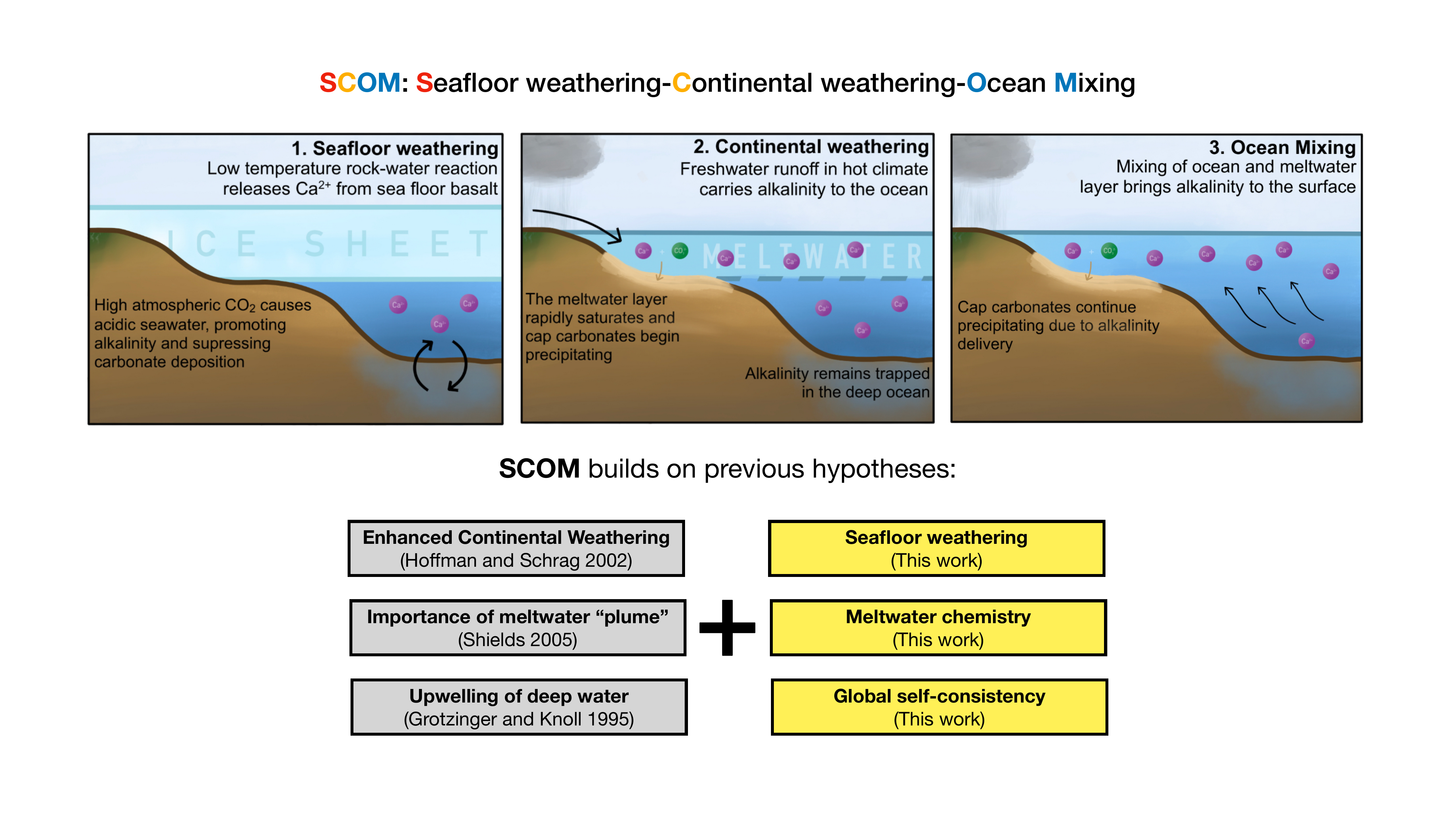}
    \caption{Summary of the SCOM mechanism and relationship to previous hypotheses.}
    \label{SCOM_summary}
\end{figure}

\begin{figure}[htbp]
    \centering
    \includegraphics[width=0.7\textwidth]{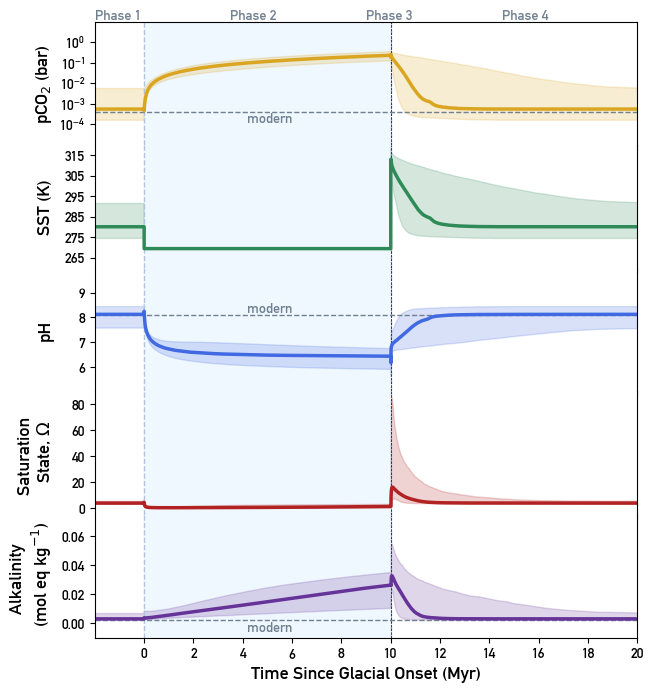}
    \caption{Climate evolution through all modeled phases with post-glacial meltwater layer duration of $t_{strat} = 10^4$ years. The shaded regions are the 95\% confidence intervals and the solid lines are the medians of 3000 model runs with randomly sampled parameters. Glaciation was terminated after 10 Myr and the stratified ocean was terminated after 10 kyr ($t_{strat} = 10^4$ years). Vertical dashed lines indicate boundaries between model phases, noting the two close vertical lines at 10 Myr and 10.01 Myr bounding the post-glacial stratified ocean phase. See Figure \ref{clim_evo_inset_supp} for a close-up on the post-glacial stratified ocean phase.}
    \label{clim_evo_supp}
\end{figure}

\begin{figure}[htbp]
    \centering
    \includegraphics[width=\textwidth]{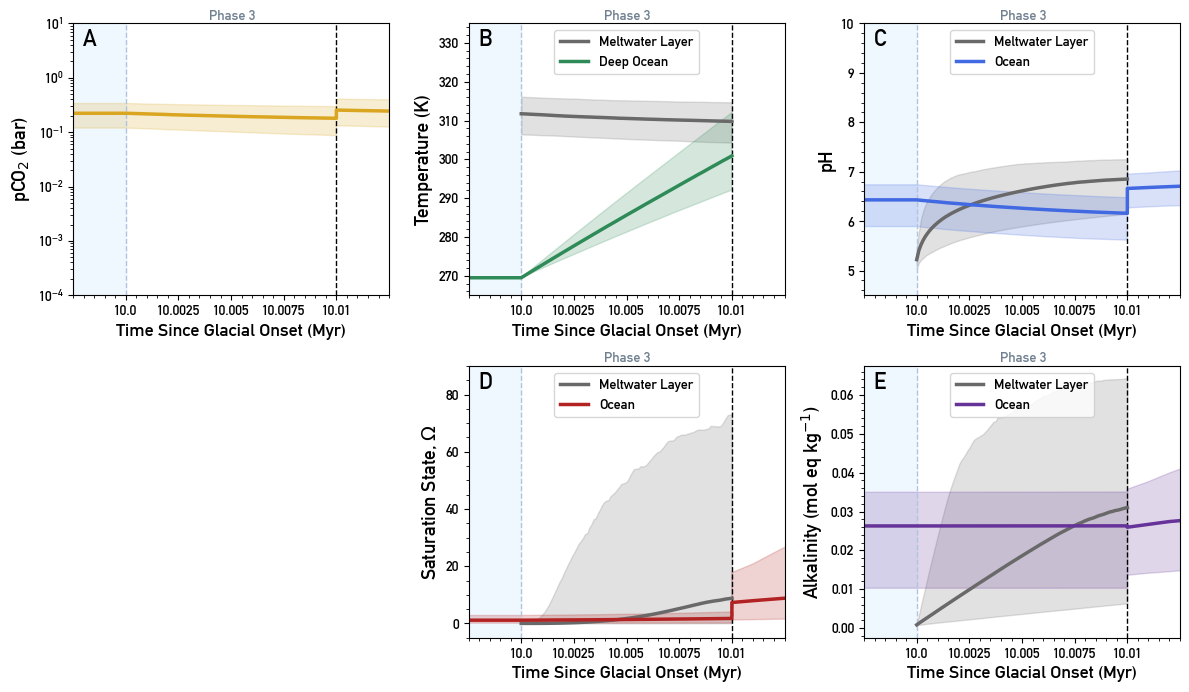}
    \caption{Climate evolution in the stratified ocean phase from Figure \ref{clim_evo_supp} with post-glacial meltwater layer duration of $t_{strat} = 10^4$ years. The shaded regions are the 95\% confidence intervals and the solid lines are the medians of 3000 model runs with randomly sampled parameters. Post-glacial ocean stratification is assumed to last 10 kyr ($t_{strat} = 10^4$ years), indicated by the vertical dashed black line. Following stratification, the two reservoirs are combined and averaged.}
    \label{clim_evo_inset_supp}
\end{figure}

\begin{figure}[htbp]
    \centering
    \includegraphics[width=\textwidth]{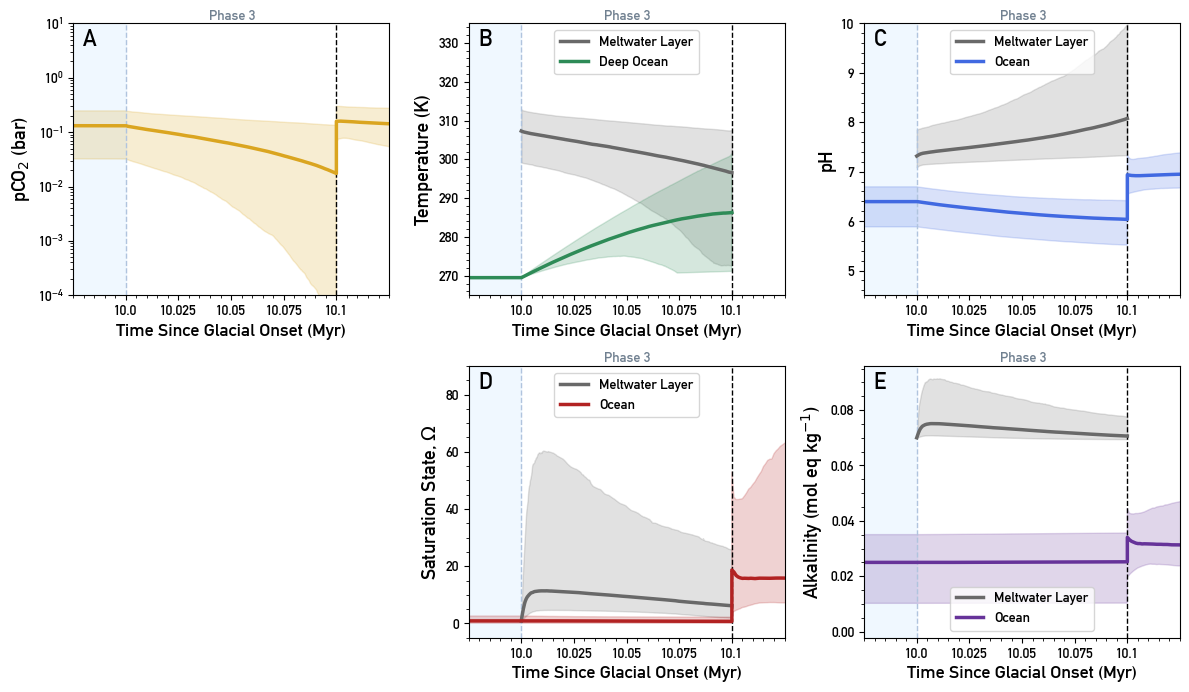}
    \caption{Climate evolution in the high-alkalinity meltwater scenario. The stratified ocean phase is shown with post-glacial meltwater layer duration of $t_{strat} = 10^5$ years. The initial alkalinity of the meltwater in phase 3 is 70,000 $\mu$mol eq kg$^{-1}$, compared to 700 - 1000 $\mu$mol eq kg$^{-1}$ in the baseline case. The shaded regions are the 95\% confidence intervals and the solid lines are the medians of 3000 model runs with randomly sampled parameters. Post-glacial ocean stratification is assumed to last 100 kyr ($t_{strat} = 10^5$ years), indicated by the vertical dashed black line. Following stratification, the two reservoirs are combined and averaged.}
    \label{clim_evo_inset_highalk}
\end{figure}


\clearpage
\section{Calculation of baseline carbon cycle fluxes, aqueous chemistry, and climate}

These calculations follow \citet{krissansen-totton_constraining_2017} and \citet{krissansen-totton_constraining_2018}. Any modifications are indicated below.

\subsection{Aqueous Chemistry}


With the carbon content and alkalinity of a given reservoir, we can calculate the other parameters of the carbon system that are used to determine fluxes and climate. To calculate the aqueous chemistry, we first define the dissolved inorganic carbon (DIC) and carbonate alkalinity (ALK):
\begin{equation} \label{chem}
\begin{split}
    {\rm DIC} & = [\ch{CO^{2-}_3}] + [\ch{HCO^-_3}] + [\ch{CO2aq}] \\
    {\rm ALK} & = 2[\ch{CO^{2-}_3}] + [\ch{HCO^-_3}] \\
\end{split}
\end{equation}

We then calculate the rest of the aqueous chemistry with the following equations \citep{pilson_introduction_2012}:
\begin{gather} \label{oceanchem}
[\ch{CO2aq}] = \ch{pCO2} \times {\rm H_{\rm CO_2}} \\
[\ch{HCO^-_3}] = \frac{[\ch{CO2aq}] \times K_1^*}{[\ch{H^+}]} \\
[\ch{CO^{2-}_3}] = \frac{[\ch{HCO^-_3}] \times K_2^*}{[\ch{H^+}]} \\
\ch{pH} = - \log_{10}([\ch{H^+}])
\end{gather}
\begin{equation} \label{s_alk}
0 = \frac{{\rm ALK}}{K^*_1 K^*_2}\left(1 + \frac{s}{\rm H_{CO_2}} \right)[{\rm H^+}]^2 + \frac{({\rm ALK - C})}{K^*_2}[\ch{H^+}] + ({\rm ALK - 2C}) \\
\end{equation}
Where H$_{\rm CO_2}$ is the Henry's law constant of \ch{CO2}, [\ch{CO2aq}] is the sum of the concentrations of dissolved \ch{CO2} and \ch{H2CO3}, and $K^*_1$ and $K^*_2$ are the first and second apparent dissociation constants of carbonic acid, respectively. The derivation of equation \ref{s_alk} and the temperature-dependent values for the dissociation constants are found in \citet{krissansen-totton_constraining_2017}. These equations are solved separately in each reservoir at each timestep. For reservoirs that are not in direct equilibrium with the atmosphere (i.e., the pore-space), $s=0$.

The carbonate saturation states of the ocean and pore-space ($\Omega_o$ and $\Omega_p$, respectively) are calculated via the equation
\begin{equation}
    \Omega = \frac{[\ch{Ca^{2+}}][\ch{CO^{2-}_3}]}{K_{\rm sp}(T)}
\end{equation}
where the concentrations and temperatures are taken individually from each reservoir. The expressions for the temperature dependent solubility product, $K_{\rm sp}(T)$, are found in \citet{krissansen-totton_constraining_2017}. 

To calculate the calcium concentration in each reservoir, we assume that all changes in carbonate alkalinity are balanced by changes in calcium concentration:
\begin{equation}
    [\ch{Ca^{2+}}] = \frac{1}{2}({\rm ALK} - {\rm ALK}_{\rm modern}) + [\ch{Ca^{2+}}]_{\rm modern}
\end{equation}
This equation effectively assumes that all changes in alkalinity are due to changes in the calcium concentration and that no other processes affect the calcium concentration. We use the modern Earth as a reference point for this approach and assume that $[\ch{Ca^{2+}}]_{\rm modern} = \SI{10.27}{mmol.kg^{-1}}$ \citep{pilson_introduction_2012}. ${\rm ALK}_{\rm modern}$ is calculated during model calibration so that the modern steady-steady state climate (e.g., surface temperature, \ch{pCO2}, and ocean pH) is reproduced. This assumption has been tested and justified for the full Earth evolution \citep{krissansen-totton_constraining_2018}, so it is suitable for calculating the background Neoproterozoic climate.

\subsection{Air and Water Temperature}

We assume that the solar luminosity in the Neoproterozoic era was 94\% of the modern value \citep{domingo_solar_1981}. To calculate Earth's surface temperature ($T_s$) as a function of \ch{pCO2}, we employ the extensively tested parameterization of a 1D radiative convective climate model used in \citet{krissansen-totton_constraining_2018}. We assume that the temperature of the ocean reservoir is equal to the atmospheric surface temperature. The temperature of the pore-space water is partially determined by the temperature of the water in the deep ocean that enters the crust. We use an empirical relationship derived in \citet{krissansen-totton_constraining_2017} to calculate the deep ocean temperature ($T_d$) as a function of $T_s$:
\begin{equation}
    T_d = a_{\rm grad}T_s + b_{\rm int}
    \label{deeptemp}
\end{equation}
where $a_{\rm grad}$ ranges from 0.8 to 1.4 and $b_{\rm int} = 274.037 - a_{\rm grad} \times 285$ K to ensure modern conditions are reproduced. The pore-space temperature, $T_p$, is related to $T_d$ via another empirical relationship identified for the Cenozoic and Mesozoic: $T_p = T_d + 9$ K \citep{coogan_alteration_2015, krissansen-totton_constraining_2017}. We assume this relationship is true because Earth's internal heat flux is not expected to have changed substantially from the modern day to the Cryogenian \citep[See][Figure 2D]{krissansen-totton_constraining_2018}. 

\subsection{Volcanic Outgassing}

We assume that the source rate of \ch{CO2} from volcanic outgassing is equal to the modern value, $6.5-10.5$ $\unit{Tmol.yr^{-1}}$ \citep{catling_atmospheric_2017}. This assumption is justified because interior evolution models indicate that Earth's internal heat flux should not have changed substantially from the modern day to the Cryogenian \citep[See][Figure 2D]{krissansen-totton_constraining_2018}. Additionally, estimates of degassing rates based on global ridge reconstructions indicate that the degassing rate at 750 Ma should have been approximately equal to the modern value \citep[See][Figure 9A]{mills_modelling_2019}.

\subsection{Continental Weathering}

As derived in \citet{walker_negative_1981}, we express continental silicate weathering as
\begin{equation}
    W_{sil} = f_{\rm w}F^{\rm mod}_{sil}\left( \frac{\ch{pCO2}}{\ch{pCO2}^{\rm mod}} \right)^{\alpha} \exp{(\Delta T_s/ T_e)}.
    \label{sil_weathering}
\end{equation}

$F^{\rm mod}_{sil}$ is the modern continental silicate weathering flux (mol eq yr$^{-1}$), calculated by running the model for modern day conditions and assuming the system is in a steady-state. Agreement between the model value and modern estimates is shown in our modern Earth benchmark. The exponent $\alpha$ is an empirical constant that determines the weathering dependence on the model atmospheric \ch{pCO2} value relative to the modern value ($\ch{pCO2}^{\rm mod}$). $\Delta T_s = T_s - T^{\rm mod}_s$ is the difference in the model global average surface temperature ($T_s)$ and the preindustrial modern value ($T^{\rm mod}_s = 285$ K), and $T_e$ is the e-folding temperature that determines the temperature dependence of the weathering rate. We explore a range of values based on existing literature estimates for the empirical constants that determine the \ch{pCO2} and temperature dependencies: $\alpha = 0.1-0.5$ and $T_e = 10 - 40$ K \citep{walker_negative_1981, volk_feedbacks_1987,sleep_carbon_2001,krissansen-totton_constraining_2017, krissansen-totton_constraining_2018}. $f_{\rm w}$ is the continental weatherability factor, described below.

Continental carbonate weathering is expressed similarly as
\begin{equation}
    W_{carb} = f_{\rm w}F^{\rm mod}_{carb}\left( \frac{\ch{pCO2}}{\ch{pCO2}^{\rm mod}} \right)^{\xi} \exp{(\Delta T_s/ T_e)}.
\end{equation}
$F^{\rm mod}_{carb} = 7-\SI{14}{Tmol.C.yr^{-1}}$ is the modern continental carbonate weathering flux, for which we explore the full range of values \citep{hartmann_global_2009}. The exponent $\xi$ determines the carbonate weathering dependence on the model $\ch{pCO2}$ and is analogous to $\alpha$ in the continental silicate weathering parameterization. We similarly explore the range $\xi = 0.1-0.5$ \citep{berner_phanerozoic_2004}. 

A key free parameter in the model is $f_{\rm w}$, the continental weatherability factor. $f_{\rm w}$ is multiplied with both the silicate and carbonate weathering rates to account for inherent uncertainties that arise when extrapolating back to the Precambrian. On one hand, landmasses with recently-rifted margins were probably concentrated at the equator throughout the Neoproterozoic as the supercontinent Rodinia broke up \citep{li_neoproterozoic_2013}. Some studies have argued that this should lead to enhanced continental weathering and a coolor background climate \citep[e.g.,][]{rooney_re-os_2014,cox_continental_2016,macdonald_initiation_2017}, but at least one carbon cycle model shows that it may have had little impact \citep{mills_modelling_2019}. On the other hand, the lack of widespread biology likely had an important and complex impact on weathering rates. Vegetation can impact weathering rates via several processes, including modifying groundwater hydrology, modifying the surface area of weatherable rocks, and modifying the water-rock contact time. Similarly, microorganisms alter the surface chemistry of weatherable rocks and potentially degrade them directly. Estimates for the biological enhancement due to weathering (i.e., biotic weathering rate divided by abiotic weathering rate) range from 0.1 to 8, with most experiments suggesting that organisms enhance weathering rather than slow it \citep[See][and references within]{wild_contribution_2022}. Because the processes described above are highly uncertain and may have offsetting effects, we group them together within the weatherability factor. We explore the range $f_{\rm w} = 0.5-1.5$ and explicitly mention when we fix $f_{\rm w}$ at a discrete value.

\subsection{Seafloor Weathering}

Seafloor weathering is the dissolution of basaltic ocean crust at low temperatures, away from mid-ocean ridges. Following \citet{krissansen-totton_constraining_2017}, we model the rate of seafloor weathering as

\begin{equation}
    W_{sea} = k_{\rm sea}\left( \frac{[\ch{H}^+]_p}{[\ch{H}^+]^{\rm mod}_p}\right)^{\gamma}\exp{(-E_{\rm bas}/RT_{\rm pore})}.
\end{equation}

Here, $k_{\rm sea}$ is a proportionality constant that reproduces the modern flux when the model is calibrated to the modern climate system, where the modern flux is approximately $\SI{0.4}{Tmol.eq.yr^{-1}}$ \citep{krissansen-totton_constraining_2018}. $[\ch{H}^+]_p$ is the model hydrogen ion molality in the pore-space and $[\ch{H}^+]^{\rm mod}_p$ is the modern value. $E_{\rm bas}$ is the effective activation energy of basalt dissolution and the exponent $\gamma$ accounts for the pH dependence of seafloor weathering. We adopt the empirically-derived ranges for these values from \citet{krissansen-totton_constraining_2017}: $E_{\rm bas} = 60-100$ kj mol$^{-1}$ and $\gamma = 0-0.5$.

We implicitly assume that other factors impacting seafloor weathering in the Neoproterozoic are negligible compared to changes in pore-space pH and temperature. For example, the seafloor weathering rate depends on geologic properties such as Earth's internal heat flux and the rate of seafloor spreading that provides fresh, weatherable basalt. As described in the above section on volcanic outgassing, interior evolution models and global ridge reconstructions indicate that these processes should not significantly impact the rate of seafloor weathering in the Neoproterozoic relative to modern \citep{krissansen-totton_constraining_2018,mills_modelling_2019}.



\subsection{Carbonate Precipitation}

On modern Earth, ocean carbonate precipitation and deposition occurs in the deep, open ocean (i.e., pelagic) and on the warmer, shallower continental shelves (i.e., neritic). Modern pelagic carbonate deposition is sourced by calcifying plankton, which did not arise until about 150 Ma, during the Jurassic period \citep{berner_phanerozoic_2004}. To our knowledge, there is no evidence for pelagic carbonates before this. Thus, we assume that carbonate deposition in the Neoproterozoic occurred only in neritic environments. 

Following \citet{ridgwell_carbonate_2003}, we parameterize the rate of continental shelf carbonate deposition as:
\begin{equation}
   P_{\rm shelf} = k_{\rm shelf}\frac{A_{\rm shelf}}{A^{\rm mod}_{\rm shelf}}(\Omega_o - 1)^n
   \label{shelf_precip}
\end{equation}
where $k_{\rm shelf}$ is a proportionality constant that reproduces the modern flux when the model is calibrated to the modern climate system. $A_{\rm shelf}/A^{\rm mod}_{\rm shelf}$ is the ratio of continental shelf area at the model time relative to the modern continental shelf area. We assume this ratio is 1 for the background Neoproterozoic climate. $\Omega_o$ is the calcite saturation state of the ocean, and $n$ determines the dependence of carbonate precipitation on the saturation state. We explore a range of $n = 1.0-2.5$ to consider the precipitation of both calcite and aragonite \citep{opdyke_carbonate_1993}. Finally, when calibrating the model to the modern climate system, we include pelagic carbonate deposition to ensure steady state balance in the same way as \citet{krissansen-totton_constraining_2017}.

The pore-space carbonate precipitation rate is treated similarly to the continental shelf rate, but there is no dependence on area:
\begin{equation}
   P_{\rm pore} = k_{\rm pore}(\Omega_p - 1)^n
\end{equation}
where $k_{\rm pore}$ is a proportionality constant that reproduces the modern flux when the model is calibrated to the modern climate system.

\clearpage
\section{Calibration and Benchmarking Against Modern and Post-glacial Earth}

\begin{figure}[htb!]
    \centering
    \includegraphics[width=\textwidth]{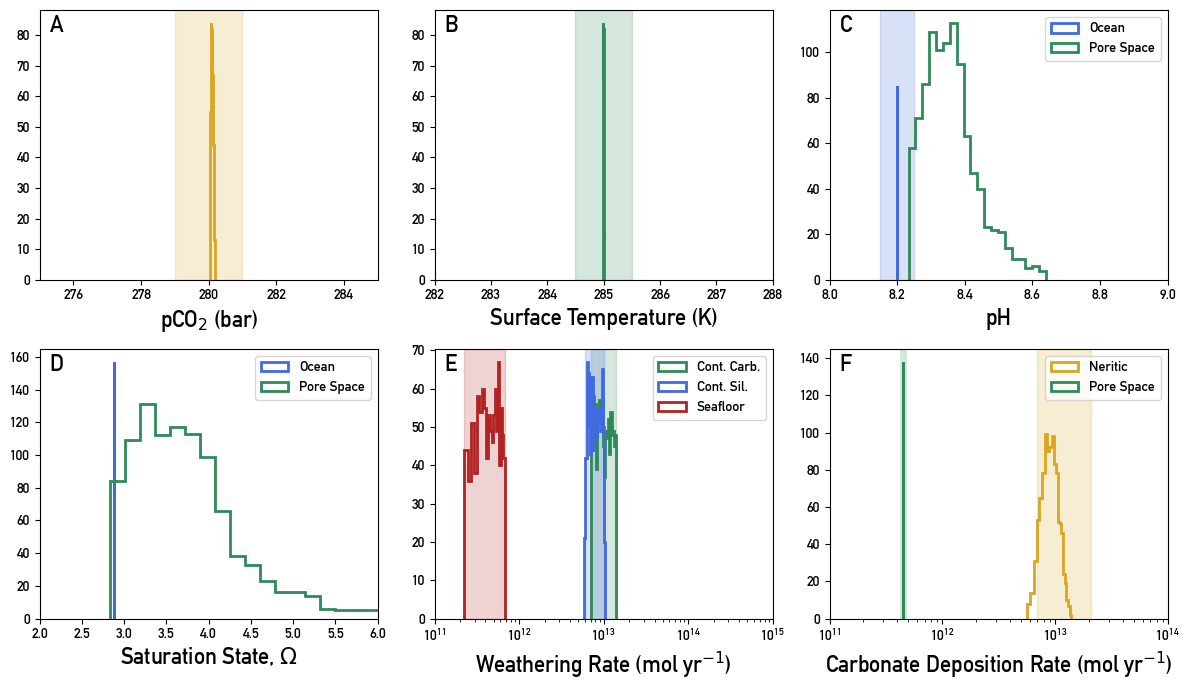}
    \caption{Model benchmark against the modern climate system. Histograms show the results for observable variables from 1000 model runs where uncertain parameters are sampled. The vertical shaded regions are the measured or assumed modern values. \textbf{(a, b, c)} We assume the pre-industrial modern climate had 280 ppm atmospheric \ch{CO2}, 285 K surface temperature, and 8.2 ocean pH. \textbf{(d)} The calcite and aragonite saturation states of the modern ocean vary from over 5 at the surface to below 0.5 in the deep ocean \citep{pilson_introduction_2012}. \textbf{(e)} Estimates of continental carbonate weathering are from \citet{hartmann_global_2009}, continental silicate weathering are from \citet{moon_new_2014}, and seafloor weathering are from \citet{coogan_evidence_2013} and \citet{krissansen-totton_constraining_2018}. \textbf{(f)} Estimates of neritic carbonate deposition are from \citet{iglesias-rodriguez_progress_2002} and pore-space deposition are from \citet{gillis_secular_2011} and \citet{krissansen-totton_constraining_2018}.}
    \label{benchmark_modern}
\end{figure}

To calibrate many of the carbon and alkalinity fluxes, we must benchmark the model against the modern system. Following \citet{krissansen-totton_constraining_2017} and \citet{krissansen-totton_constraining_2018}, we assume that the system is in a steady state at modern day, and thus we can set all of the governing differential equations equal to zero. We then enforce the measured values of the modern outgassing rate, the modern continental carbonate weathering rate, the modern pore-space carbonate precipitation rate, and the ratio of seafloor weathering to pore-space carbonate precipitation. From this, we directly calculate the required modern values of seafloor weathering, continental silicate weathering, and ocean carbonate precipitation that balance the system. This approach is justified in \citet{krissansen-totton_constraining_2017} and \citet{krissansen-totton_constraining_2018} because only the relative sizes of the carbon and alkalinity fluxes impact the calculation of observable values. A sample of model runs in the modern climate state compared to measured values is shown in Figure \ref{benchmark_modern}.

In phase 1, we run the model until a steady state is reached. We define a steady state as the point in the model evolution when \ch{CO2} would change by less than 1\% if its rate of change is projected 1 billion years into the future.

\begin{figure}[htb!]
    \centering
    \includegraphics[width=\textwidth]{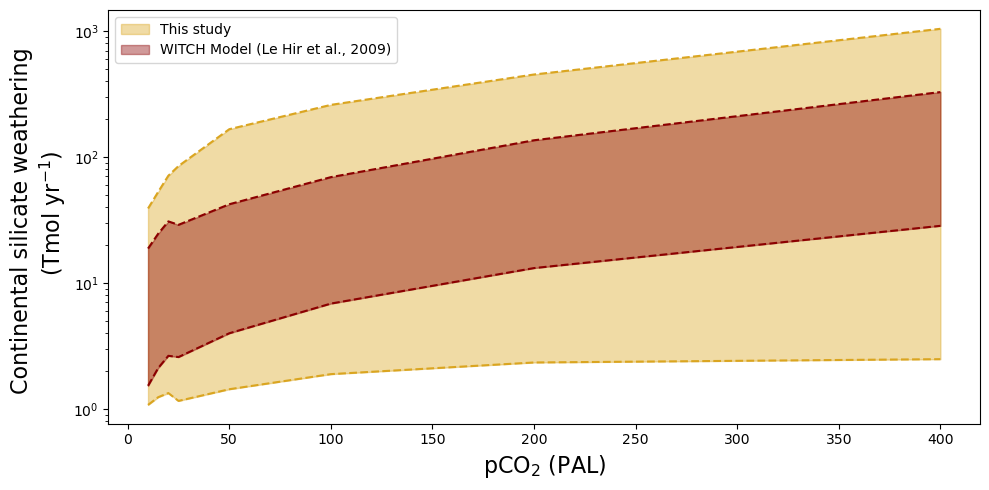}
    \caption{Post-glacial continental weathering rate predictions as a function of \ch{pCO2} from this study versus \citet{le_hir_snowball_2009}.}
    \label{postglacial_weathering}
\end{figure}

Comparison to the WITCH weathering model \citep{le_hir_snowball_2009} for post-glacial Earth is shown in Figure \ref{postglacial_weathering}

\clearpage
\section{Model Transitions}

\subsection{Glaciation}

The initiation of the glaciation is instantaneous and forced in our model. Previous coupled climate and ice-sheet models indicate that the onset of global glaciation should be rapid, on the order of 2 kyr, due to the ice-albedo feedback \citep[e.g.,][]{hyde_neoproterozoic_2000}. This is below the typical $\sim1$ Myr timescale on which the geologic carbon cycle impacts climate. Consistent with the purpose of this study, we do not model the evolution of climate and surface chemistry during this relatively short transition interval. Thus, we are effectively modeling this transition period within a single timestep.

\subsection{Deglaciation and Ocean Stratification}

When entering phase 3, we nominally assume that all ice sheets and glaciers melt instantaneously. This is consistent with the many positive feedbacks associated with melting (e.g., the ice-albedo instability) that cause predictions of rapid melting on the timescale of 2 kyr \citep{hyde_neoproterozoic_2000,hoffman_snowball_2017}. However, it is possible that melting was more prolonged than we nominally assume, on the order of 100 kyr \citep[e.g.][]{Fairchild.2023}.

Here we present a sensitivity study to evaluate how a prolonged deglaciation would impact our model results. To do this, we modify the phase 3 model configuration so that the mass of the meltwater layer ($M_s$) changes over time, due to the melting of ice sheets. To incorporate this in the model, we first have to modify the governing system of equations (Eqs 2 in main text) related to the meltwater layer. Previously, $M_s$ was assumed to be constant and could be factored out in the differential equations. Now $M_s = M_s(t)$, so it cannot be factored out and must be differentiated via the chain rule. This accounts for the addition of new water into the meltwater layer as the glaciers melt, where we assume that the carbon concentration of the added water is determined by equilibration with the atmosphere and we assume that the alkalinity of the added water is the same as our nominal model scenario. The governing equations for carbon and alkalinity in the meltwater layer in phase 3 become
\begin{equation} \label{sensitivity_system}
    \begin{split}
    \frac{dC_s}{dt} & = \frac{1}{M_s(t)}\Bigl( -K({\rm DIC_s} - C_d) + V_{\rm air} + W_{\rm carb} - P_{\rm shelf} - O_{\rm sink} \Bigr) - C_s\frac{M_s'(t)}{M_s(t)} \\
    \frac{dA_s}{dt} & = \frac{1}{M_s(t)}\Bigl( -K(A_s - A_d) + 2W_{\rm sil} + 2W_{\rm carb} - 2P_{\rm shelf} \Bigr) + (A_{\rm melt} - A_s)\frac{M_s'(t)}{M_s(t)} \\
    \end{split}
\end{equation}
where $M_s'(t) = \frac{dM_s(t)}{dt}$ and $A_{\rm melt}$ is the alkalinity of the meltwater that is being added to the layer.

We now consider 2 scenarios for $M_s(t)$ in the prolonged deglaciation scenario. In both cases, we assume it takes 100 kyrs for all ice to melt to test the most extreme endmember. First, we consider a scenario where melting occurs linearly with time:
\begin{equation}
    M_s(t) = M_s(t_{strat}) \times \left(0.1 + 0.9\frac{t}{t_{strat}}\right)
    \label{linearmelt}
\end{equation}
where $M_s(t_{strat})$ is the mass of the meltwater layer after deglaciation is complete, the same as we assume in the nominal model described in Methods. We assume the meltwater layer starts at 10\% of its final mass and then linearly increases until the final mass is reached at the same time ocean stratification ends (100 kyrs). Second, we consider a scenario where melting occurs in 3 pulses:
\begin{equation}
    M_s(t) = M_s(t_{\rm strat}) \times \left(0.1 + \frac{0.9/3}{1 + e^{-100(t/t_{\rm strat} - 0.25)}} + \frac{0.9/3}{1 + e^{-100(t/t_{\rm strat} - 0.5)}} + \frac{0.9/3}{1 + e^{-100(t/t_{\rm strat} - 0.75)}}\right).
    \label{pulsedmelt}
\end{equation}
The above equation is the sum of 3 logistic functions that approximate stepwise increases in $M_s(t)$ at 25, 50, and 75 kyrs. We again assume that the meltwater layer starts at 10\% of its final mass.

We also use the melt percentage ($M\% = M_s(t)/M_s(\rm final)$) to scale the continental weathering rates and sea surface temperature. First, we simply multiply the continental weathering rates by $M\%$ to approximate the reduction in weathering due to coverage of weatherable land by ice sheets and the slow hydrologic cycle. Second, our climate model does not take surface albedo as an input, so we approximate the changing albedo of the surface by using $M\%$ to linearly scale the surface temperature ($T_s(t)$) between the assumed snowball ocean temperature (269.5 K) and the result of the climate model ($T_{\rm s, clim}(\rm pCO_2)$):
\begin{equation}
    T_s(t) = 269.5 + M\%(T_{\rm s, clim}(\rm pCO_2) - 269.5).
    \label{melttemp}
\end{equation}

\begin{figure}[htbp]
    \centering
    \includegraphics[width=\textwidth]{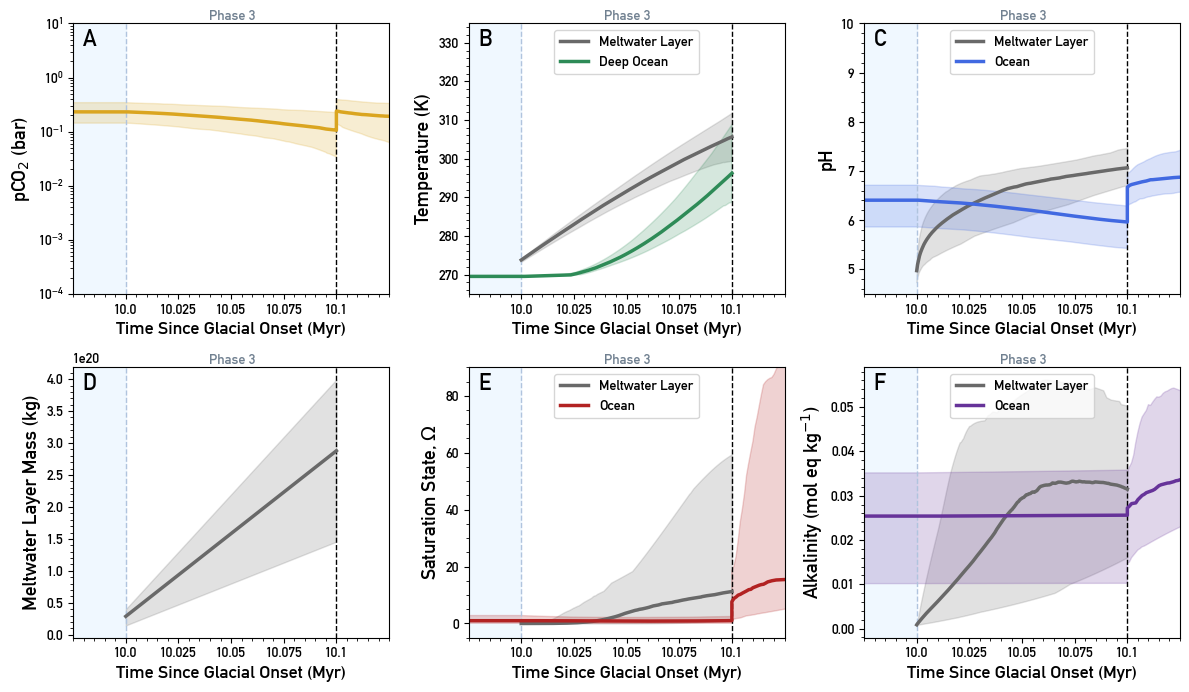}
    \caption{Climate evolution in sensitivity study with linear deglacation. The shaded regions are the 95\% confidence intervals and the solid lines are the medians of 3000 model runs with randomly sampled parameters.}
    \label{linear_evol}
\end{figure}

\begin{figure}[htbp]
    \centering
    \includegraphics[width=\textwidth]{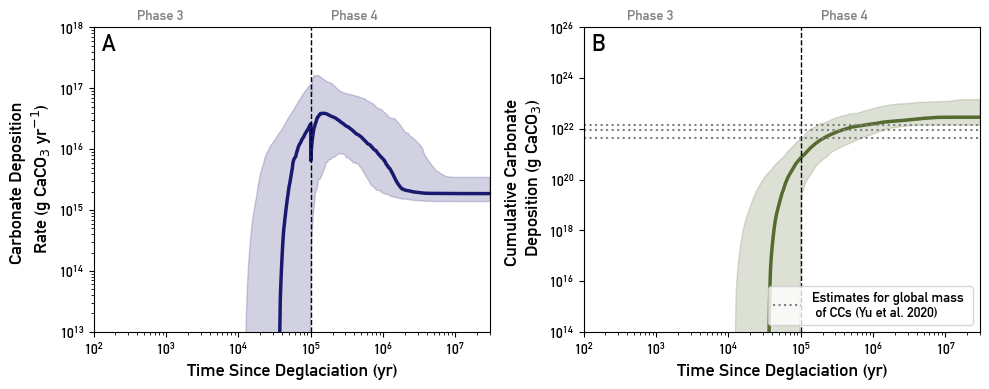}
    \caption{Carbonate deposition rate and cumulative carbonate deposition in the sensitivity study with linear deglaciation, corresponding to Figure \ref{linear_evol}. The shaded regions are the 95\% confidence intervals and the solid lines are the median model runs with randomly sampled parameters.}
    \label{linear_carbs}
\end{figure}

\begin{figure}[htbp]
    \centering
    \includegraphics[width=\textwidth]{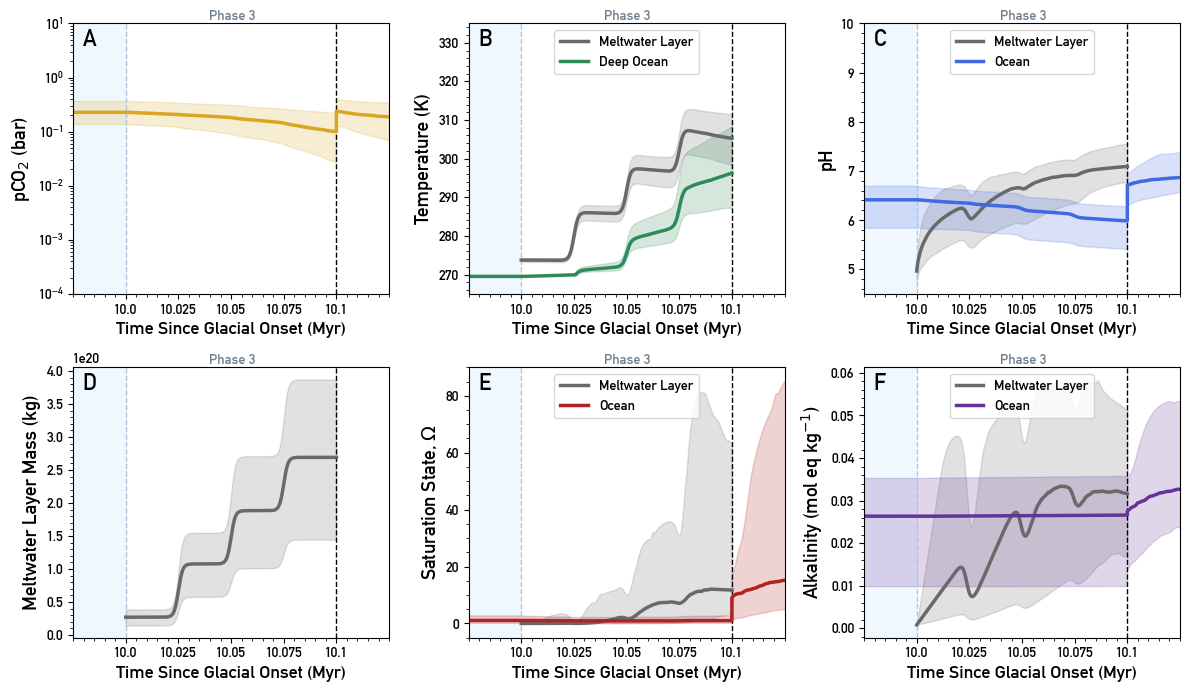}
    \caption{Climate evolution in sensitivity study with pulsed deglacation. The shaded regions are the 95\% confidence intervals and the solid lines are the medians of 3000 model runs with randomly sampled parameters.}
    \label{pulsed_evol}
\end{figure}

\begin{figure}[htbp]
    \centering
    \includegraphics[width=\textwidth]{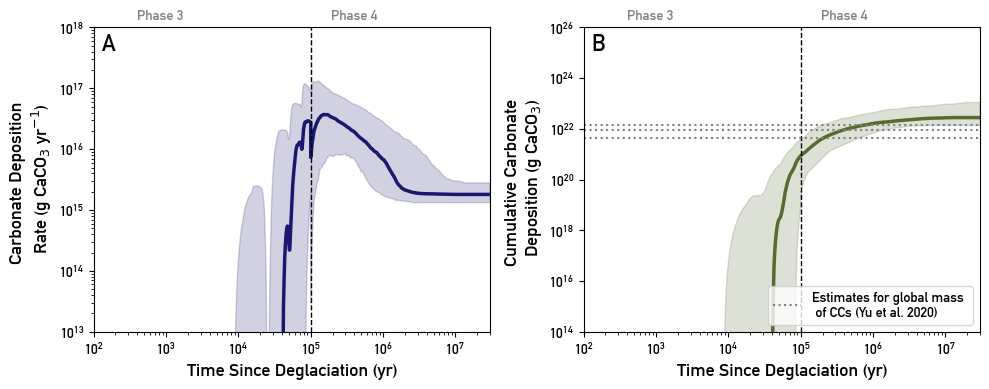}
    \caption{Carbonate deposition rate and cumulative carbonate deposition in the sensitivity study with pulsed deglaciation, corresponding to Figure \ref{pulsed_evol}. The shaded regions are the 95\% confidence intervals and the solid lines are the median model runs with randomly sampled parameters.}
    \label{pulsed_carbs}
\end{figure}

The results of the sensitivity study with linear glaciation are shown in Figures \ref{linear_evol} and \ref{linear_carbs}, and with pulsed deglaciation are shown in Figures \ref{pulsed_evol} and \ref{pulsed_carbs}. 
The mass of the meltwater layer can have a noticable impact on the meltwater chemistry. This is especially visible for the pulsed deglaciation case where other climate variables reflect the stepwise increases in meltwater mass. This leads to differences in the rate of CC deposition. In the nominal model, the minimum global mass of CCs is deposited in 32 kyr to 591 kyr. In the linear deglaciation sensitivity study, the minimum global mass of CCs is deposited in 101 kyr to 682 kyr. In the pulsed deglaciation sensitivity study, the minimum global mass of CCs is deposited in 109 kyr to 618 kyr.

Prolonged deglaciation in the nominal sensitivity study increases the depositional timescale by up to 90 kyr. This is because of the reduced meltwater volume and the reduction in weathering rates and thus alkalinity delivery to the meltwater layer, which is required for CC deposition. However, the sensitivity study shown here is the ''worst case'' endmember scenario because (1) we assume a very long deglaciation time of 100 Myr (2) we assume weathering rates and surface temperature are reduced in direct proportion to the amount of deglaciation, which is likely an overestimate, (3) we assume that the added meltwater has very low alkalinity, and (4) we assume that none of the alkalinity from the deep ocean mixes into the meltwater layer, although this would likely occur given the smaller meltwater mass and lower flux to the ocean. If any of these assumptions are relaxed, the depositional timescale will likely approach our nominal model results. Additionally, because we were forced to make simple modifications to our existing weathering rates and climate model to approximate the partial coverage of ice sheets, a more detailed weathering study in the future would help to further investigate this topic.

\subsection{Ocean Mixing}

When phase 3 ends, we assume that the ocean instantaneously overturns and becomes well-mixed. The ocean dynamics models predict that this process should be more gradual, with slow mixing that eventually overturns the ocean by the time $t_{\rm strat}$ is reached \citep{yang_persistence_2017,ramme_climate_2022}. Thus, we implicitly overestimate the intensity of stratification for a given value of $t_{\rm strat}$; however, we test two values for $t_{\rm strat}$ to mitigate this. We maintain a simplified approach here to avoid additional model uncertainty during this transition and to focus on the more prominent geochemical trends during phase 3.

